\documentclass{aa}

\usepackage{graphicx}
\usepackage{txfonts}
\usepackage{natbib}
\usepackage{graphicx}
\usepackage{amssymb}
\usepackage{amsmath}%for the split
\usepackage{subcaption}
\begin{document}

    \title{Detectability of continuous gravitational waves from isolated neutron stars in the Milky Way: the population synthesis approach}
    %TODO
    \titlerunning{GWs from synthetic, isolated NSs in the Milky Way}

   \author{Marek Cie\'{s}lar \inst{1}\fnmsep\thanks{mcie@camk.edu.pl}
          \and
          Tomasz Bulik \inst{2}\fnmsep\thanks{tb@astrouw.edu.pl}
          \and
          Ma{\l}gorzata Cury{\l}o \inst{2}
          \and
          Magdalena Sieniawska \inst{2,1}
          \and
          Neha Singh \inst{2}
          \and
          Micha{\l} Bejger \inst{1}
          }
   \authorrunning{Cie\'{s}lar}

   \institute{Nicolaus Copernicus Astronomical Center, Polish Academy of Sciences, Bartycka 18, 00-716 Warsaw, Poland
         \and
         Astronomical Observatory, University of Warsaw, Al. Ujazdowskie 4, 00-478 Warsaw, Poland
         }

   %\date{Received September 15, 1996; accepted March 16, 1997}

% \abstract{}{}{}{}{} 
% 5 {} token are mandatory
 
  \abstract
  % context heading (optional)
  % {} leave it empty if necessary  
   {}
  % aims heading (mandatory)
   {We estimate the number of pulsars, detectable as continuous gravitational wave sources with the current and future gravitational-wave detectors, assuming a simple phenomenological model of evolving non-axisymmetry of the rotating neutron star.
   }
  % methods heading (mandatory)
   {We employ a numerical model of the Galactic neutron star population, with the properties established by comparison with radio observations of isolated Galactic pulsars. We generate an arbitrarily large synthetic 
   population of neutron stars and evolve their period, magnetic field, and position in space. 
   We use a gravitational wave emission model based on exponentially 
   decaying ellipticity -- a non-axisymmetry of the star, with no assumption of the origin of a given 
   ellipticity. We calculate the expected signal in a given detector for a $1$ year 
   observations and assume a detection criterion of the signal-to-noise ratio of $11.4$ -- comparable to a targeted continous wave search.
   We analyze the population detectable separately in each detector:  
   Advanced LIGO, Advanced Virgo, and the planned Einstein Telescope. In the calculation of the expected signal we neglect signals frequency change due to the source spindown and the Earth motion with respect to the Solar barycentre.}
  % results heading (mandatory)
   {With conservative values for the neutron stars evolution: supernova rate once 
   per $100$ years, initial ellipticity $\epsilon_0 \simeq 10^{-5}$ with no 
   decay of the ellipticity $\eta = t_{\rm{hub}} \simeq 10^{4}\,\rm{Myr}$, the expected number of detected neutron stars is below one: $0.15$ (based on a simulation of $10\,\rm{M}$ stars) for the Advanced LIGO detector. A broader study of the parameter space ($\epsilon_0$, 
   $\eta$) is presented. With the planned sensitivity for the Einstein Telescope,
   and assuming the same ellipiticity model, the expected detection number is: $26.4$ pulsars during a $1$-year long observing run.}
  % conclusions heading (optional), leave it empty if necessary 
   {}


   \maketitle
%
%________________________________________________________________
\section{Introduction}
%\citep{Acernese_2014} \citep{AdvLIGO2015}
The discovery of a double black hole merger \citep{PhysRevLett.116.061102} and a 
binary neutron star (NS) merger \citep{PhysRevLett.119.161101,2017ApJ...848L..12A} %Komentarz Magdy o cytowaniu - ale czy to nie jest styl AA?
initiated an era of observable gravitational waves (GW) sources. The first catalogue of transient GW events \citep{PhysRevX.9.031040}, created after two observational runs O1 and O2, contains $11$ sources and the observational run O3 brought nearly $1$ statistically-significant detection candidate per week\footnote{\url{https://gracedb.ligo.org/superevents/public/O3/}} \citep{2020arXiv201014527A}. 
Still, some sources of the 
gravitational radiation, in particular the continuous waves from rotating, deformed (non-axisymmetric) 
NSs, elude the detection and we only have upper limits on their intrinsic asymmetry (\citet{PhysRevD.93.042007}, \citet{PhysRevD.96.122004}, and \citet{2019PhRvD.100b4004A}, \citet{2020arXiv200912260S}, \citet{2020PhRvL.125q1101D}).
The community actively pursues many novel methods of detection 
\citep[see][]{1998PhRvD..58f3001J,PhysRevD.70.082001,Antonucci_2008,PhysRevD.90.042002,PhysRevD.98.102004,2e829abe62dc4158b2b0e895b23c2c1d,Astone_2010,Piccinni_2018,PhysRevD.72.102002,PhysRevD.98.084058,2019PhRvD.100f4013C,2019PhRvL.123j1101D,2020PhRvD.102b2005D}.
%\citep[see][]{2012arXiv1201.3176P,2015GReGr..47...11A,2017MPLA...3230035R,}.
Proposed continuous GW emission models include the r-modes  \citep{1998PhRvL..80.4843L,Bildsten_1998, PhysRevD.58.084020, Andersson_1999} and a fixed rotating axis-asymmetric anisotropy
of the quadrupole moment of inertia \citep{PhysRevD.20.351, 1996A&A...312..675B}. 
In this work, we  concentrate only on the latter case. 
For more in-depth review of the current state of GW emission models from NSs, see \citet{2011GReGr..43..409A, 2015PASA...32...34L, 2017MPLA...3230035R, 2019Univ....5..217S}.

\subsection{Aim of the research}
A general estimate of the detectability of a population of NSs in the gravitational 
emission regime was previously investigated by \citet{2000A&A...359..242R}, \citet{2005MNRAS.359.1150P}, \citet{PhysRevD.78.044031}, and \citet{Woan_2018}. 
We revisit this problem by providing a
realistic approach to the properties of the  estimated population of isolated (solitary, not in a binary system) NSs in our Galaxy,
and by more detailed treatment of the detection of the signal. We limit ourselves 
to a phenomenological parametrization of the deformation of a NS, to find the general constraints on models without investigating physical details that drive the asymmetry of a star.
We aim to estimate the  fraction of the Galactic population of NSs that
can be detected by the Advanced LIGO \citep{AdvLIGO2015}, the Advanced Virgo \citep{2015CQGra..32b4001A},
and by the Einstein Telescope \citep{2010CQGra..27s4002P,Maggiore_2020}. We assume a coherent signal intergration time to be $t_{\rm{obs}} = 1$ year.  Although such 
period of time is comparable to the current O3 observation run of the LIGO-Virgo
 collaboration, it should be noted that the usable, scientific data does not cover the whole period (due to maintenance, quality assurance, etc.). Thus, such an estimate is the optimistic upper limit of the possible detection rate.

\section{The physical model}

As a basis for our studies we use the numerical models developed in our previous work 
\citep{2020MNRAS.492.4043C}. We simulated a population of $10^7$ isolated NSs\footnote{$10^7$ NSs is roughly $20$ times more then an expected Galactical population. This number represent a trade-off between statistical error, and computational speed.}, with properties similar to the observed, single radio pulsars in the Galaxy \citep{2005AJ....129.1993M}\footnote{\url{http://www.atnf.csiro.au/research/pulsar/psrcat/}}. 
In our population synthesis model, we include the evolution of period, period derivative, as well as magnetic field of each NS and trace its position as it moves in the Galactic potential.
The  magnetic field decay influences the evolution of rotation period and period  derivative assuming the magnetic dipole model. 
We propagate
the stars through the Galactic potential, from their birth places (inside the spiral
arms), with inclusion of supernova kicks \citep{2005MNRAS.360..974H}, to their final positions at
the current date, when they are observed. 

\subsection{Kinematics}
In our NS model we simulate the initial position distribution of newly born NSs in spiral arms of the Galaxy and we propagate the stars through the Galactic gravitational potential. Since each NS is born during a supernova explosion, we draw the {\em initial velocity} from the \citet{2005MNRAS.360..974H} model.
\subsection{Period and magnetic field evolution}

\subsection{Radio luminosity}
In our model we simulate two phenomenological radio luminosity models. One is proportional to the {\em rotational} energy of the NS:
\begin{equation}
    L_{\rm{rot}}\sim \left(\dot{P}_{15}^{\frac{1}{3}}P^{-1}\right)^{\kappa_{1}}
\end{equation}
And the other is a {\em power-law}
    \begin{equation}
    L_{\rm{p-l}}\sim P^{\kappa_{2}}\dot{P_{15}}^{\kappa_{3}}
\end{equation}
where $\kappa_{i}$ are the parameters of the models\footnote{Symbols describing parameters of the luminosity models where simplified in regards to our previous work \citep{2020MNRAS.492.4043C}}, and $\dot{P}_{15}$ is a period derivative divided by $10^{-15}$.
Since both models produce a population that is identical to the observed one, choosing one over another does not have any implication on the evolution of NS in the context of this work. Therefore we use the parameters from the {\em rotational} model as it contains less parameters. 

\subsubsection{The age and size of the simulated population}
The maximum age of a simulated NS is $t_{ \rm{age},\rm{max} }=50\,\rm{Myr}$, while the number of stars equals to $N_{\rm{pop}}=10^{7}$. With an assumption that NS is born every $100\,\rm{yr}$, this equals to $20$ times higher synthetic population in the last $50\,\rm{Myr}$ period. The factor $20$ was picked due to the technical aspects of the simulation -- providing significant sample to compute statistics whereas limiting the simulation time.

The exclusion of NSs older than $50\,\rm{Myr}$ is dictated by the fact that due to the exponential magnetic decay, implemented in the model, the NSs evolve faster in the period-period derivative plane. While considering their {\em characteristic age} $\tau = P/2\dot{P}$, our population is comparable to the observed one and reaches to characteristic ages of $10\,\rm{Gyr}$. Moreover, due to the fact, that older, solitary NS reach a region of period-period derivative plane known as the {\em graveyard} (a region of periods longer than $1s$) and stay there due to the very low period derivative ($\dot{P}<10^{-22}\,[\rm{s/s}]$). This region is outside of the frequency range of the current detectors (Advanced LIGO and Advanced Virgo) as well as planned, next generation observatories (Einstein Telescope).

\subsubsection{Model parameters}
The parameters of the model used in this paper are shown in Table \ref{tab:params}. The distributions describing the population's parameters and details of the pulsar evolution model are described in  \citet{2020MNRAS.492.4043C}. The pulsar population model provides us with a population of NSs, and for each of them we have its position in the sky, distance, rotation period, period derivative, and the pulsar age.

\begin{table}                                                                    
\centering                                                                       
\begin{tabular}{lr}                                                              
Parameter & Value \\ 
\hline                                                                           
$\log {\left(\Delta/{\rm Myr}\right)}$  &  $0.63$\\ 
$\log \left( \widehat{B}_{{\rm init}}/{\rm G} \right)$ &  $12.67$\\ 
$\log \left( \sigma_{B_{{\rm init}}}/{\rm G} \right)$  &  $0.34$\\ 
$\widehat{P}_{{\rm init}}\,{\rm s}$ &  $0.05$\\ 
$\sigma_{P_{{\rm init}}}\,{\rm s}$  &  $0.07$\\
\hline                                                                           
\end{tabular}                                                                    
\caption{The values of the parameters describing the NS evolution model: $\Delta$ is the magnetic field decay timescale, $\widehat{B}_{\rm{init}}$ and $\sigma_{B_{\rm{init}}}$ are the mean initial magnetic field and it dispersion (log-normal distribution), and $\widehat{P}_{init}$ and $\sigma_{P_{init}}$ are the mean initial rotation period and its dispersion (positive only, normal distribution).}                                                       
\label{tab:params}                                                                 
\end{table}  

\subsection{Millisecond Pulsars exclusion from the Model}
\label{sec:MSPModel}
A standard model for of evolution of Millisecond Pulsars (MSPs) indicate that they are produced due to the accretion in a Low Mass X-ray Binaries \citep[see][]{1991PhR...203....1B}, although, it should be noted that not all observed MSPs can be confidently described with such model \citep{2009ApJ...693L.109K}. 
Modelling a standard MSP model with an accretion phase requires a very detailed treatment of the binary interactions \citep[e.g.][]{2008ApJS..174..223B}. 
This treatment of binary interactions is beyond scope of our NS evolution model presented in out previous work \citep{2020MNRAS.492.4043C}, thus we limit this work to the probing of only solitary NSs. 
%We discuss how the result might change due to the inclusion of MSPs in the Sec. \ref{sec:MSPDisc}. 

\subsection{Gravitational wave signal from a rotating ellipsoid}
The GW's signal registered in the detector is given by the dimensionless amplitude:
\begin{equation}
h = F_{+} h_{+} + F_{\times} h_{\times}.
\end{equation}
where the $F_+$ and $F_\times$ (described in detail in Sec. \ref{seq:AntennaPattern}) denote the antenna pattern functions. Since the two wave's polarisations ($+$ and $\times$) are orthogonal, the measured quantity is the root of the square of the $h$:
\begin{equation}
    \sqrt{h^{2}} = \sqrt{\left(F_{+} h_{+} + F_{\times} h_{\times}\right)^{2}}.
\end{equation}
We assume a general model of a rotating triaxial ellipsoid after \citet{1996A&A...312..675B}.
We do not assume the cause of the deformation, but merely state that it
is necessary for the time-varying quadruple moment of the NS mass (for different theoretical models of NS deformation see \citet{1969ApJ...157.1395O}, \citet{1969Natur.224..781M}, \citet{1970Natur.228..655C}, and \citet{1972ARA&A..10..335P}). The NS's ellipticity is defined as:
\begin{equation}
    \epsilon = \frac{\left| I_{1} - I_{2} \right|}{I},
\end{equation}
where $I$ is the mean NS moment of inertia with respect to the rotational axis, and 
$I_{1}$ and $I_{2}$ are moments in the  plane perpendicular to the principal  axis. The GW's amplitudes in both polarisations can be written as:
\begin{equation}
\begin{split}
  h_{+} & = h_{0} \sin \chi \left( A \cos \left( \frac{2\pi(t-t_{0})}{P} \right)  -B  \cos \left( \frac{2\pi(t-t_{0})}{P/2} \right) \right), \\  
h_{\times} & = h_{0} \sin \chi   \left(C \sin \left( \frac{2\pi(t-t_{0})}{P} \right)  -D  \sin \left( \frac{2\pi(t-t_{0})}{P/2} \right) \right),
\end{split}
\label{eq:hplushcross}
\end{equation}
where $A  = \frac{1}{2} \cos \chi \sin \iota \cos \iota$, $B = \sin \chi \frac{1+\cos^{2}\iota}{2}$, $C = \frac{1}{2} \cos \chi \sin \iota$, and $D = \sin \chi \cos \iota$. The NS signal is parametrised by: the angle between the rotation axis and the major axis of the elipticity $\chi$, the line of sight inclination $\iota$, the rotational period $P$, the time $t$, and the time corresponding to a phase-shift $t_{0}$. The amplitude $h_{0}$ is defined as: 
\begin{equation}
h_{0} = \frac{16\pi^{2}G}{c^4} \frac{I \epsilon}{P^{2} r},
\label{eq:h0full}
\end{equation}
where $G$ is the gravitational constant, $c$ is the speed of light, $r$ distance to the star.

\subsubsection{The average emission over the NS Period}
Due to paremeters $A$, $B$, $C$, and $D$ being idependent of the time, we compute an average of $h^{2}$ over the period $P=1/\nu$, where $\nu$ is the frequency: 
\begin{equation}
\begin{split}
\left< h^{2} \right>_{P}  =& \left< ( F_{+} h_{+} + F_\times h_\times )^{2} \right> = \\
=& h_{0}^{2} \sin^{2}\chi
\left< \left(F_{+} A \cos2\pi\nu t + F_{+} B \cos 2\pi 2\nu t\right.\right. \\ 
&\left.\left. + F_{\times} C \sin2\pi\nu t + F_{\times} D \sin 2\pi 2\nu t\right)^{2} \right> = \\
 =& \frac{1}{2} h_0^{2} \sin^{2}\chi \left(F_{+}^{2} \left(A^{2}+B^{2}\right) + F_{\times}^{2}\left(C^{2}+D^{2}\right)\right).
\end{split}
\end{equation}
The term $\left< h^2 \right>_P$ represents the average power of emitted GW signal at two harmonics: $\nu=\frac{1}{P}$ and $2\nu=\frac{2}{P}$. 
In our analysis we emulate the simplest approach of detecting a NS at a single frequency with its amplitude above a certain level (see Seq. \ref{seq:SNR}). Therefore, we treat the amplitude of the GW in each frequency independently:
\begin{equation}
    \begin{split}
        \left< h^{2} \right>_{P,\nu} &=  \frac{1}{2} h_0^{2} \sin^{2}\chi \left(F_{+}^{2} A^{2} + F_{\times}^{2}C^{2}\right), \\
        \left< h^{2} \right>_{P,2\nu} &=  \frac{1}{2} h_0^{2} \sin^{2}\chi \left(F_{+}^{2} B^{2} + F_{\times}^{2}D^{2}\right).
    \end{split}
    \label{eq:hhPfreq2freq}
\end{equation}

\subsubsection{The antenna pattern and the sidereal day average}
\label{seq:AntennaPattern}
The detectors are described by four parameters. Their geographical position longitude and 
latitude, the angle between their arms, and the angle measured counter-clockwise between the East and the bisector of the arms.
The NS location in the sky is described by the equatorial coordinates: the right ascension and the declination, as well as the polarisation angle (the orientation of the rotation axis with respect to the line of sight).
%#all input angles should be in radians
%#const double az_rad, #theat, the azimuth 
%#const double alt_rad,#phi, the altitude
%#const double psi,   #the polarisation angle
%#const double beta,  #the latitude of the detector
%#const double lambda,# the longitude of the detector
%#const double eta,   #the angle between the arms of the detector
%#const double chi    #the bisector of the arms points in this direction#
We follow the work of \citet[][see Eqs. 10-13 therein]{1998PhRvD..58f3001J} with the antenna power in $+$ and $\times$ wave's polarisations. We average the antenna pattern over one full rotation of the Earth, yielding a final average (over the NS period and a sidereal day). The square of registered GW amplitude is:
\begin{equation}
\left< h^2 \right>_{\rm{P,D}} = \frac{1}{2} h_0^2 \sin^2\chi\, \left(\left< F_+^2 \right>(A^2+B^2)+
\left< F_\times^2 \right>(C^2+D^2)   \right).
\end{equation}
Similar to the Eq. \ref{eq:hhPfreq2freq}, we divide the period-, day-average into two separate harmonics: 
\begin{equation}
    \begin{split}
        \left< h^{2} \right>_{P,D,\nu} &=  \frac{1}{2} h_0^{2} \sin^{2}\chi \left(\left<F_{+}^{2}\right> A^{2} + \left<F_{\times}^{2}\right>C^{2}\right), \\
        \left< h^{2} \right>_{P,D,2\nu} &=  \frac{1}{2} h_0^{2} \sin^{2}\chi \left(\left<F_{+}^{2}\right> B^{2} + \left<F_{\times}^{2}\right>D^{2}\right). \\
    \end{split}
    \label{eq:hhPDfreq2freq}
\end{equation}	

\subsection{The detectors sensitivity and the signal integration}\label{seq:SNR}
For the Advanced LIGO and Advanced Virgo detectors, we assume the amplitude spectral density of the noise $\sqrt{S_h(\nu)}$ (expressed in units of $\rm{strain}/\sqrt{\rm{Hz}}$)  from the first three-months of the O3 observation run\footnote{For the sensitivity curves see \url{https://dcc.ligo.org/LIGO-T2000012/public}.} to be representative and constant for the whole period of our analysis \citep[][living review, see arXiv version for the latest update]{2018LRR....21....3A}.  
We also analyse the future Einstein Telescope (ET-D) in the D configuration (three detection arm pairs in a shape of the equilateral triangle, see \citet{Hild_2011}). The location of this facility has not been chosen yet. In that regard, we assume the same location for ET-D as for the Advanced Virgo detector.
We scale the sensitivity\footnote{For the sensitivity curve see \url{https://tds.virgo-gw.eu/?content=3&r=14065}.} appropriately for the D configuration by a factor $1/(\sqrt{3}sin(\pi/3))$ \citep[see][]{Hild_2011}.
Also, we sum the signal from its three arms (E1, E2, E3) as $\left< h^2 \right>_{\rm{P,D}}^{\rm{ET-D}} = \left< h^2 \right>_{\rm{P,D}}^{\rm{E1}} + \left< h^2 \right>_{\rm{P,D}}^{\rm{E1}} + \left< h^2 \right>_{\rm{P,D}}^{\rm{E3}}$.

Since the signal from a NS comes at two harmonics ($\nu$ and $2\nu$, see Eq. \ref{eq:hhPfreq2freq}), we define the signal-to-noise (SNR) ratio \citep[see][ch. 7]{jaranowski_analysis_2009} separately for each harmonic:
\begin{equation}
    \rm{SNR}_{\nu} = \frac{ \sqrt{\left< h^{2}_{P,D,\nu} \right>}\sqrt{t_{\rm{obs}}}}{\sqrt{S_h(\nu)}},\,\,\,\,
    \rm{SNR}_{2\nu} = \frac{ \sqrt{\left< h^{2}_{P,D,2\nu} \right>}\sqrt{t_{\rm{obs}}}}{\sqrt{S_h(2\nu)}},
\end{equation}
where $t_{\rm{obs}}$ is the integration time. The $t_{\rm{obs}}$ is far greater then period $P$ or a sidereal day $D$ over which the signal from NS is averaged (see Eq. \ref{eq:hhPDfreq2freq}). We assume a detection threshold for a pulsar of $\rm{SNR}_{\nu,2\nu}=11.4$, and the integration time is one year $t_{\rm{obs}}=31536000$ seconds.
Thus we use the detectability threshold
\begin{equation}
    h_{0,\nu}^{*} \sim 11.4 \sqrt{\frac{S_{h}(\nu)}{t_{\rm{obs}}}}, \\
    h_{0,2\nu}^{*} \sim 11.4 \sqrt{\frac{S_{h}(2\nu)}{t_{\rm{obs}}}},
\end{equation}
and state that a NS is detected in a given harmonic (either $\nu$ or $2\nu$) is 
the averaged, detected strain crosses the threshold:
\begin{equation}
    \sqrt{\left< h^{2}_{P,D,\nu} \right>} \geq h_{0,\nu}^{*}, \\
    \sqrt{\left< h^{2}_{P,D,2\nu} \right>} \geq h_{0,2\nu}^{*}. 
\end{equation}

The value for the SNR threshold used in other searches varies from $5$ to almost $12$. For the O2 blind search \citep{2019PhRvD.100b4004A} it's equal to $5.3$. For both low and high frequency bands in O1 blind search \citep{2018PhRvD..97j2003A, 2017PhRvD..96f2002A} the value is equal to $5$. In transient signal searches \citep{2018CQGra..35f5009A} as well in the narrow-band searches \citep{2019PhRvD..99l2002A} the SNR/threshold is equal to $8$. For the single a single-template search for the known pulsars it's equal to $11.4$ after \citet{PhysRevD.98.084058}:
\begin{equation}
    h_{0}^{*} \sim 11.4 \sqrt{\frac{S_{h}}{T_{\rm{obs}}}}
\end{equation}

We justify using the value of $11.4$ due to the fact that it is most conservative as well as our simulation resembles a single-template search of a known pulsar -- we assume that we know the position in the sky and neglect the error associated with it, as well as we compare the signal with a corresponding frequency bin in the detector noise.

\subsection{Neutron star asymmetry model}

Neutron star must be asymmetric with respect to the rotation axis  to emit GWs. There is a plethora of models that predict some level of asymmetry. In this work we do not want to concentrate on any particular physical model. Rather than that we propose a simple, two parameter, phenomenological model of evolution of a NS asymmetry. We assume that there is an initial asymmetry $\epsilon_0 $ at the time the NS is formed and it decays on a timescale of $\eta$.
 Thus, at a given moment of time the asymmetry is given by: 
 \begin{equation}
     \epsilon = \epsilon_0 \exp\left(\frac{-t}{\eta}\right).
 \end{equation}
With such a model we can explore a general parameter space of possible mechanisms of asymmetry generation and place observational constraints in such space. 
 
\section{Results and discussion}

We can now proceed to the estimation of the detectability of the GW form NSs. 
We  use the standard model of properties of NS population \citep{2020MNRAS.492.4043C} and for each NS we calculate the expected GW signal-to-noise ratio given its actual position in the sky, distance, period and ellipticity. For each model of evolution of ellipticity characterized by two parameters $\epsilon_{0}$ and $\eta$ we  calculate the expected number of observed NSs. We normalize our results to a population where a NS is formed in the Milky Way every 100 years, however we estimate the  population for a much larger model population ($10\,\rm{M}$ stars, $1$ every $5\rm{yr}$). Thus, after normalization we can obtain the expected number of observed NSs below unity. 

\begin{figure*}
\centering
\begin{tabular}{cc}
\begin{subfigure}{0.5\textwidth}\centering
    \includegraphics[width=1\columnwidth]{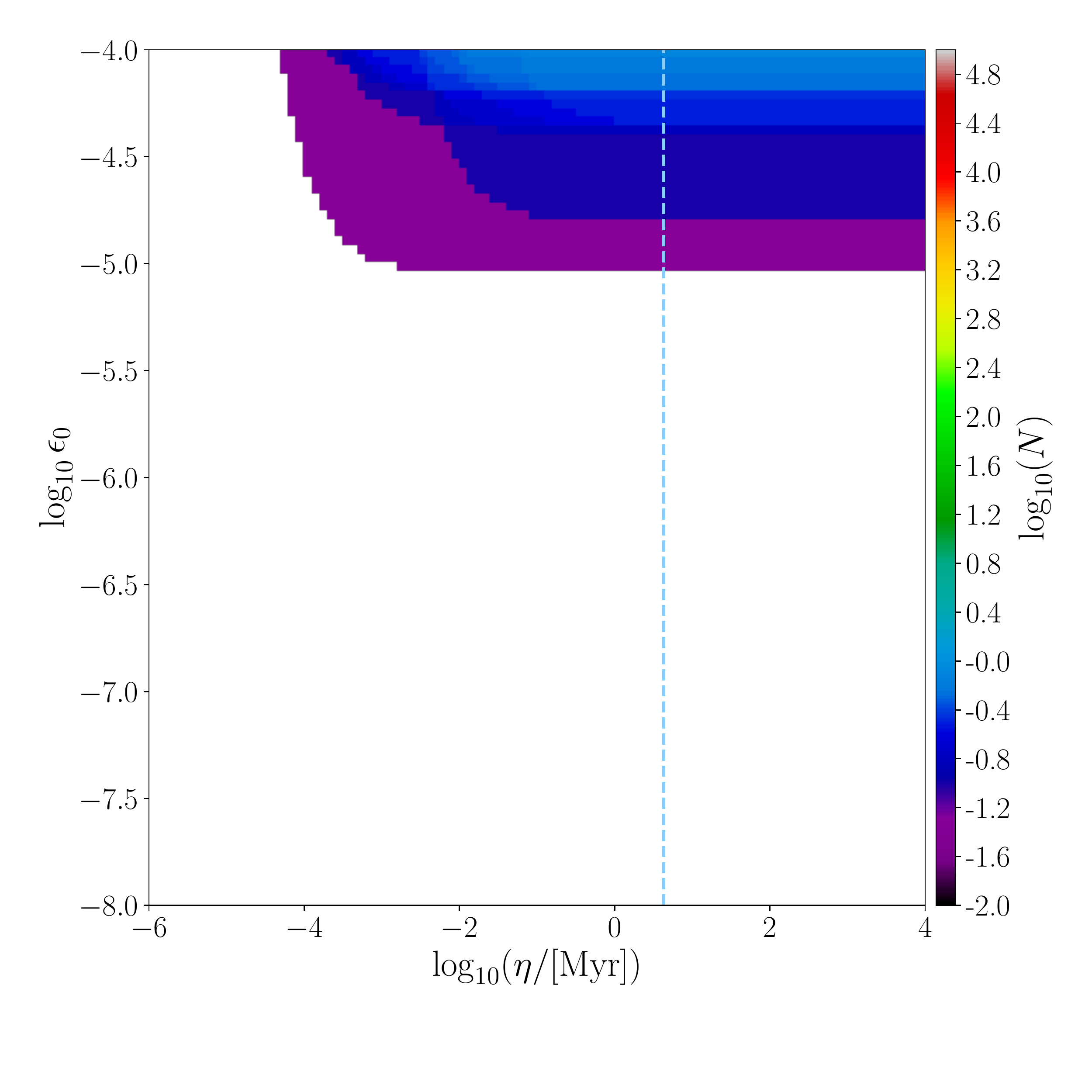}
    \vspace{-2em}
    \caption{$\nu$. L1 Advanced LIGO detector at Livingston.}
    \label{fig:EtaEpsilonL1}
\end{subfigure}
&
\begin{subfigure}{0.5\textwidth}\centering
    \includegraphics[width=1\columnwidth]{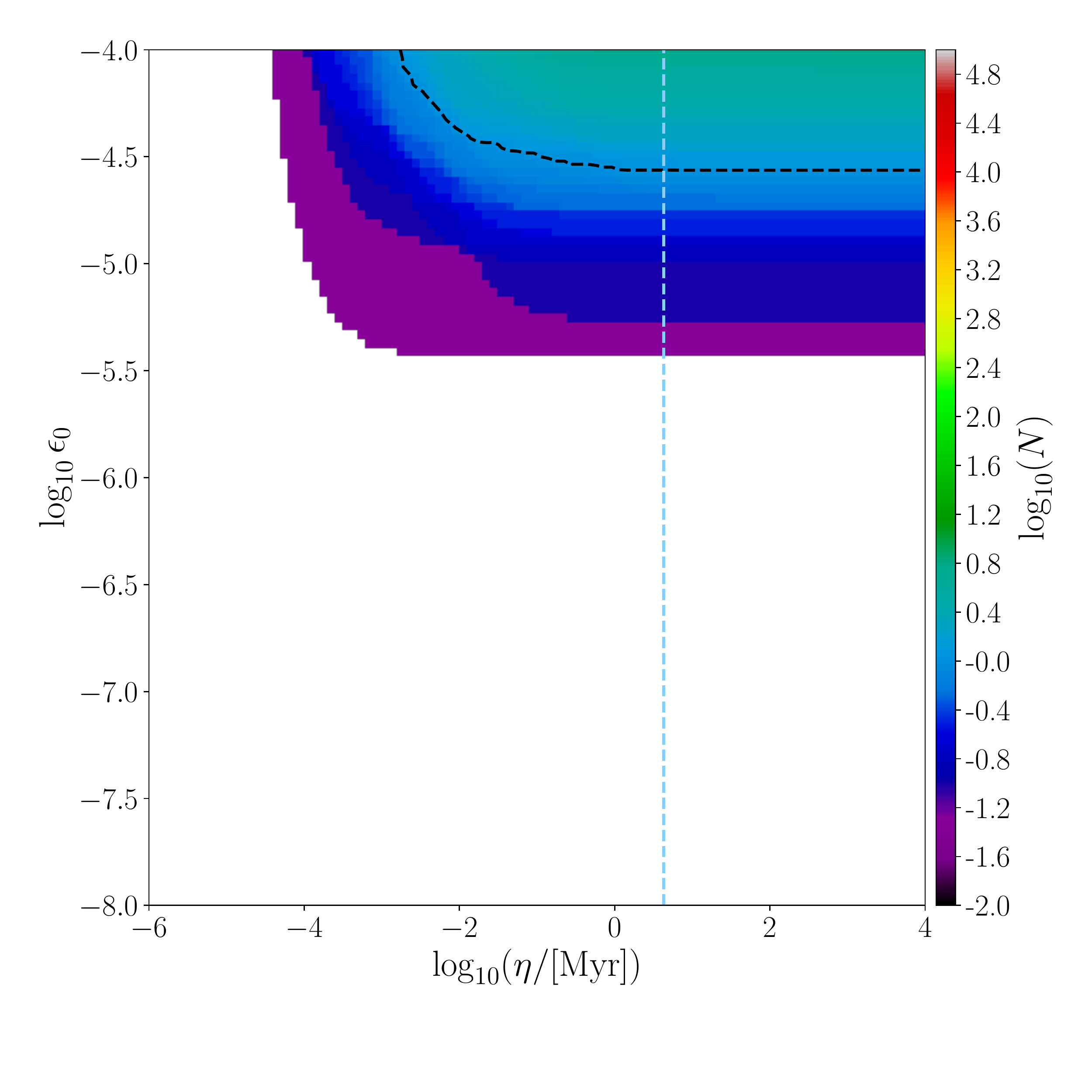}
    \vspace{-2em}
    \caption{$2\nu$. L1 Advanced LIGO detector at Livingston.}
    \label{fig:EtaEpsilonL1_05}
\end{subfigure}
\\
\begin{subfigure}{0.5\textwidth}\centering
    \includegraphics[width=1\columnwidth]{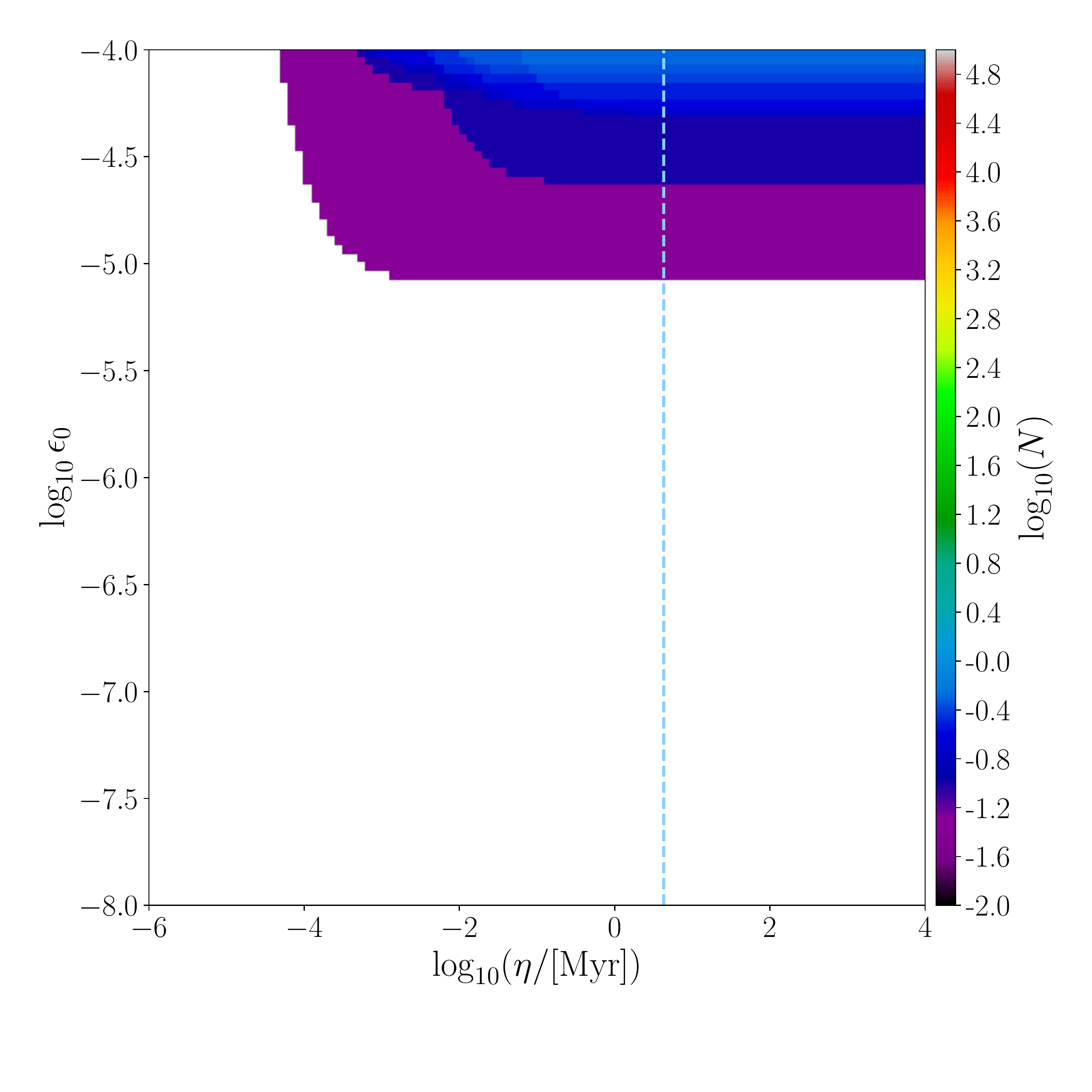}
    \vspace{-2em}
    \caption{$\nu$. H1 Advanced LIGO detector at Hanford.}
    \label{fig:EtaEpsilonH1}
\end{subfigure}
&
\begin{subfigure}{0.5\textwidth}\centering
    \includegraphics[width=1\columnwidth]{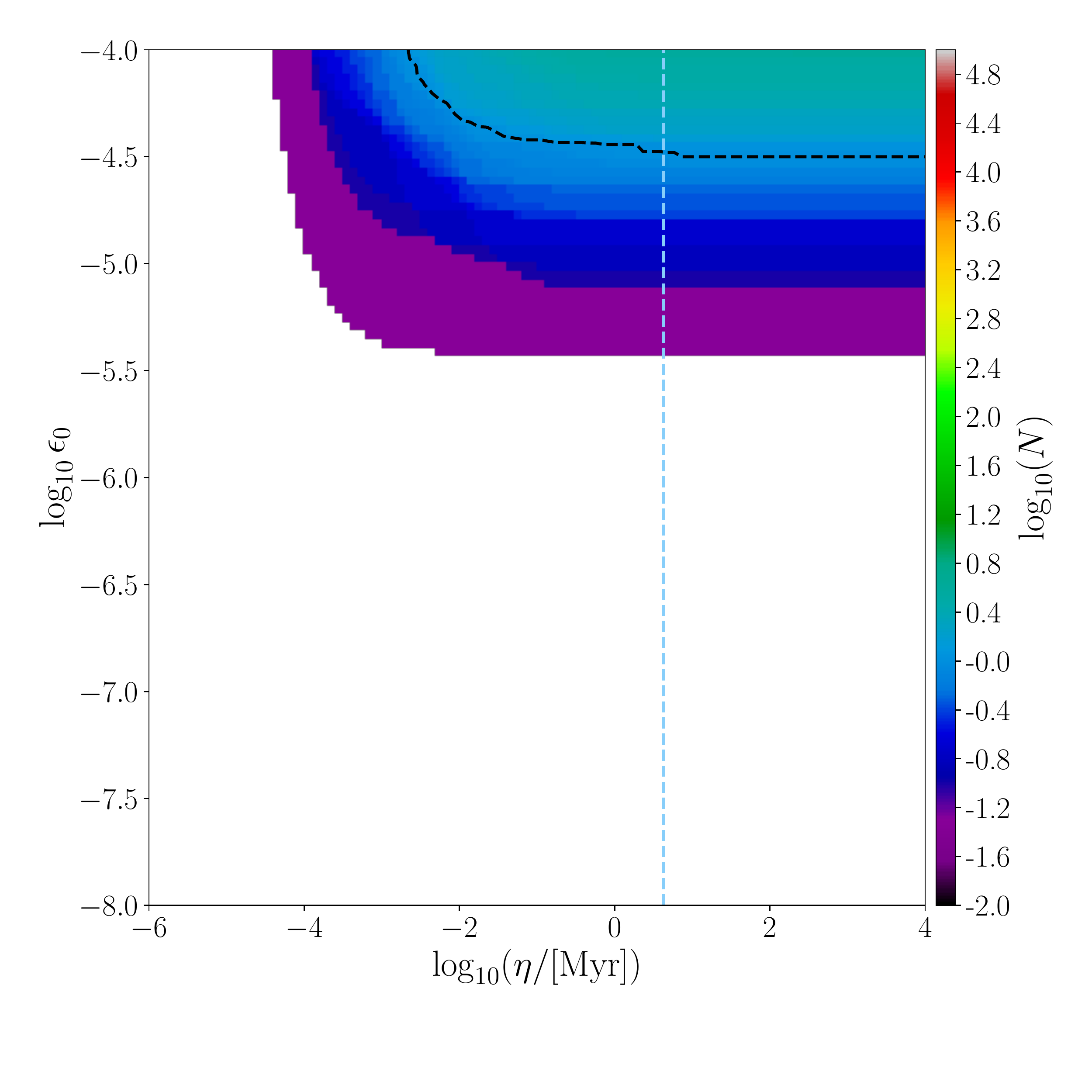}
    \vspace{-2em}
    \caption{$2\nu$. H1 Advanced LIGO detector at Hanford.}
    \label{fig:EtaEpsilonH1_05}
\end{subfigure}

\\
\end{tabular}
\caption{The number of observable pulsars in one year observations in the space of the parameters of the model $\eta-\epsilon_{0}$ for the Advanced LIGO detectors. The left column corresponds to the signal's $\nu$ harmonic, the right column the $2\nu$ harmonic. The color represents the expected number of detection for each model. The black dashed line in each plot corresponds to the models where we expect to see one pulsar. All models below the black dashed line correspond to less than one expected detection. The blue, vertical, dashed line indicates models where $\eta$ is equal to the magnetic field decay $\Delta$ (see Tab. \ref{tab:params}).}
\label{fig:eta_epsilon_multifigure_A}
\end{figure*}

We present the results for the Advanced LIGO Livingston in Figs.  \ref{fig:EtaEpsilonL1} and \ref{fig:EtaEpsilonL1_05} and for the Advanced Virgo detector in  Figs. \ref{fig:EtaEpsilonV1} and \ref{fig:EtaEpsilonV1_05} for the signals' $\nu$ and $2\nu$ harmonics respectively (see Eq. \ref{eq:hhPfreq2freq}). The color in these Figs. corresponds to the number of detectable NSs in the entire sky for each model as a function of the parameters $\epsilon_{0}$ and $\eta$. The dark dashed line in each plot corresponds to the models with the expected one detection, and we expect less than one detection in the region of parameter space below this line. A very low number of stars crossing the detection criteria in some regions of the plots may induce arbitrary shapes due to a too low static (most prominently seen on in Figs. \ref{fig:EtaEpsilonED_05} and \ref{fig:EtaEpsilonED}). We disregard such shapes as a significant increase of simulated stars would be needed to smooth the low-detection region in Figs. \ref{fig:eta_epsilon_multifigure_A} and \ref{fig:eta_epsilon_multifigure_B}.
The detectability of pulsar population with Advanced detectors with the noise level factor two lower (below 200Hz) then the current one, could improve the detectability by factor of 3 for the H1 detector and factor 7 for the L1 detector. Though, such increase would remain below a single detected NS.

For the models where the population evolves slowly, i.e. $\eta > 0.1\,\rm{Myr}$, we expect at least one detection, for the $2\nu$ harmonic, with the Advanced LIGO detectors provided that the initial  ellipticity $\epsilon_{0} > 2.5\times 10^{-5}$ and $\epsilon_{0} > 6.3\times 10^{-5}$ for the Advanced Virgo detector. For the population of quickly evolving ellipticity the detectability quickly becomes more and more difficult as the timescale $\eta$ decreases. This is due to the fact the number of NSs with sufficiently large ellipiticity in the Milky Way at a given time becomes smaller and smaller. 

The similar diagrams calculated  for the sensitivity of ET are shown in Figs.  \ref{fig:EtaEpsilonED} and \ref{fig:EtaEpsilonED_05}. For this detector in the regime of slowly varying ellipticity, i.e. with $\eta > 0.1\,\rm{Myr}$ the detectable models have the initial ellipticity $\epsilon_{0} > 7.9\times 10^{-7}$ for the $2\nu$ harmonic, and $\epsilon_{0} > 5.6\times 10^{-6}$ for the $\nu$ harmonic. The population becomes undetectable for $\eta<100\,\rm{yr}$, as in this case the ellipticity decreases on the timescale comparable to the time between consecutive supernovae explosions in the Galaxy.

\begin{figure*}
\centering
\begin{tabular}{cc}

\begin{subfigure}{0.5\textwidth}\centering
    \includegraphics[width=1\columnwidth]{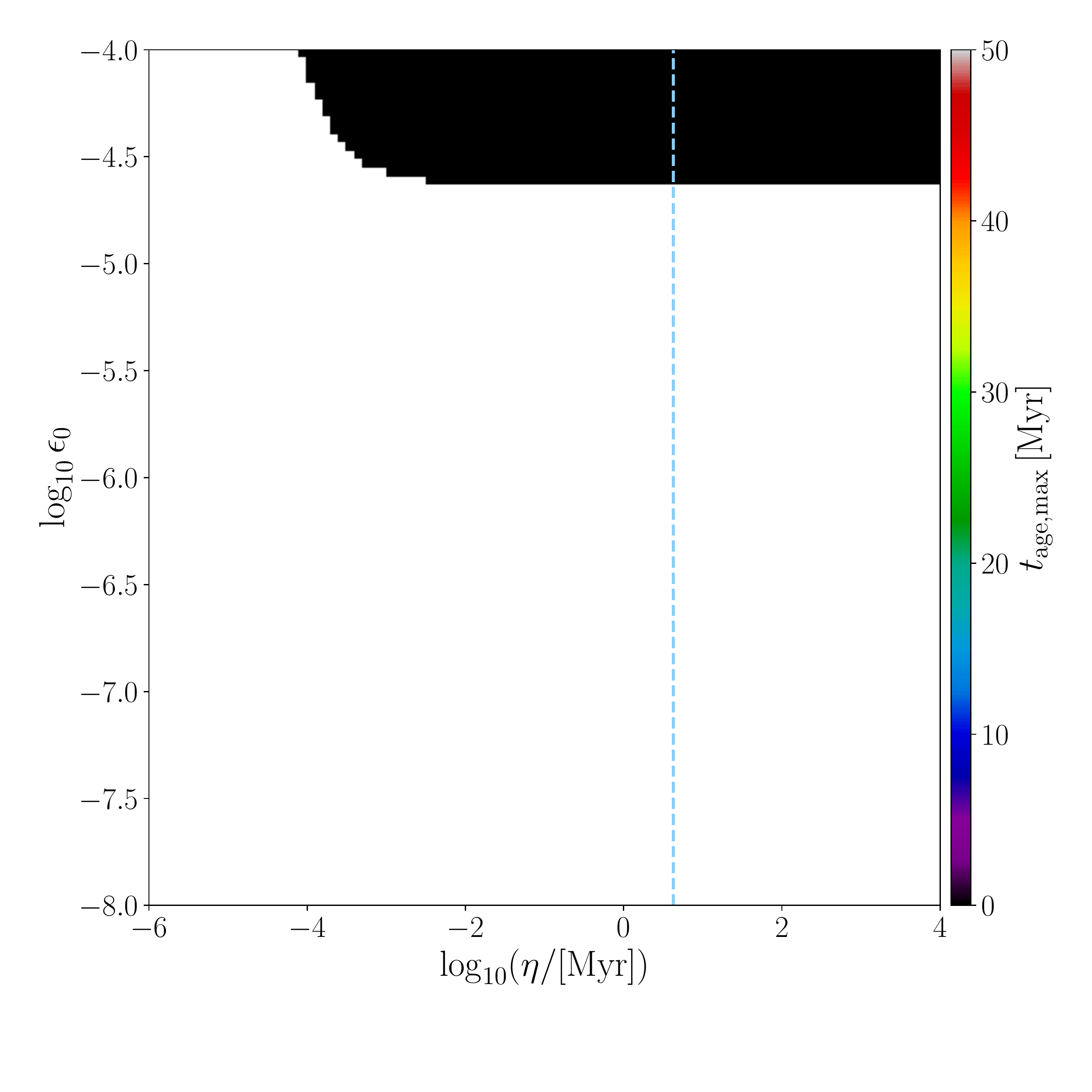}
    \vspace{-2em}
    \caption{$\nu$. V1 Advanced Virgo detector.}
    \label{fig:MaxAgeVirgo}
\end{subfigure}
&
\begin{subfigure}{0.5\textwidth}\centering
    \includegraphics[width=1\columnwidth]{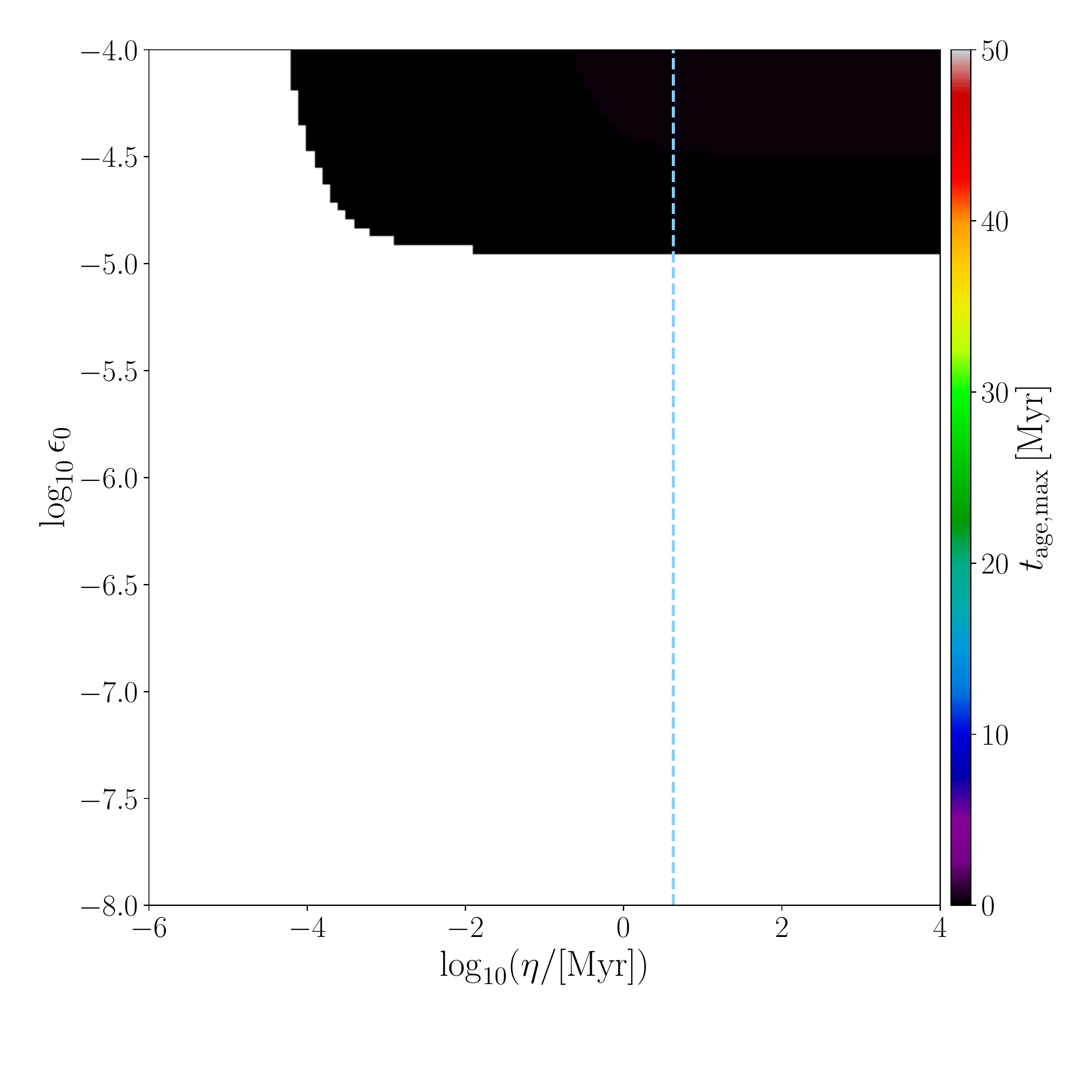}
    \vspace{-2em}
    \caption{$2\nu$. V1 Advanced Virgo detector.}
    \label{fig:MaxAgeVirgo_05}
\end{subfigure}
\\
\begin{subfigure}{0.5\textwidth}\centering
    \includegraphics[width=1\columnwidth]{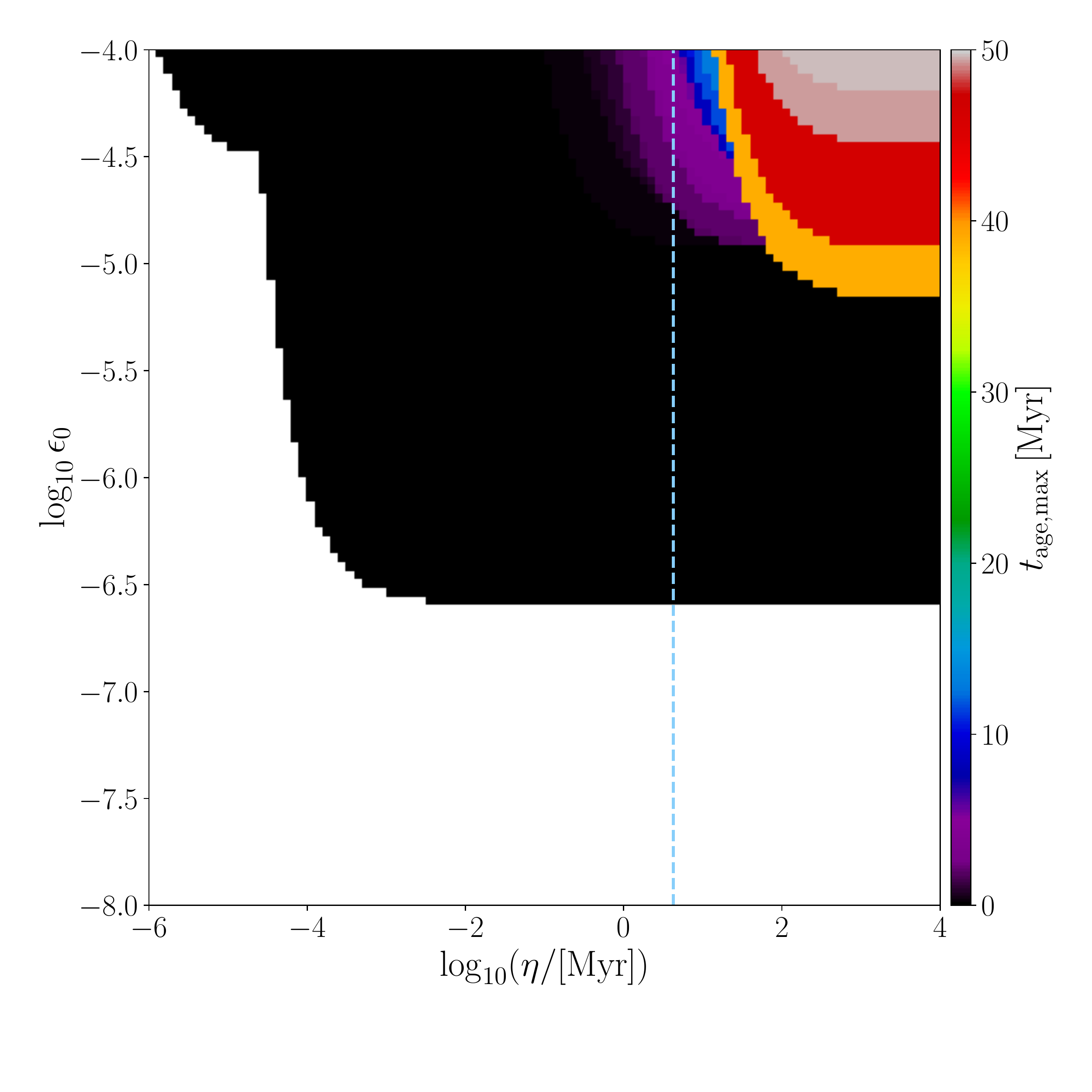}
    \vspace{-2em}
    \caption{$\nu$. Einstein Telescope detector, configuration D.}
    \label{fig:MaxAgeED}
\end{subfigure}
&
\begin{subfigure}{0.5\textwidth}\centering
    \includegraphics[width=1\columnwidth]{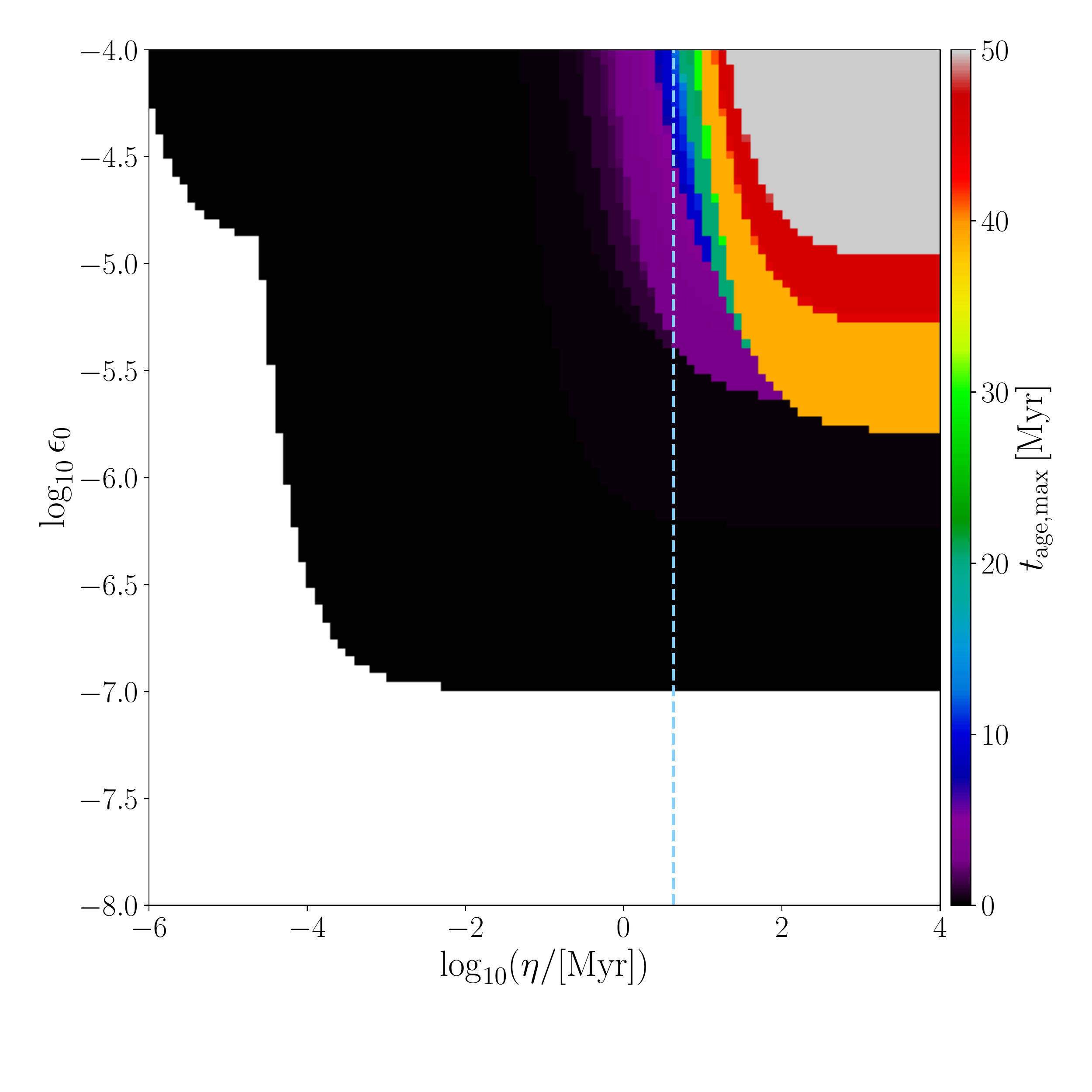}
    \vspace{-2em}
    \caption{$2\nu$. Einstein Telescope detector, configuration D.}
    \label{fig:MaxAgeED_05}
\end{subfigure}
\\
\end{tabular}
\caption{The maximum age of visible NSs in one year observations in the space of the parameters of the model $\eta-\epsilon_{0}$ for the Advanced Virgo, and Einstein Telescope detectors. The left column corresponds to the signal's $\nu$ harmonic, the right column the $2\nu$ harmonic. The color represents the maximum age in the population for a given model. The blue, vertical, dashed line indicates models where $\eta$ is equal to the magnetic field decay $\Delta$ (see Tab. \ref{tab:params}).}
\label{fig:max_age_multifigure_B}
\end{figure*}

It is quite interesting to investigate in more detail the properties of the detectable population. 
In Fig. \ref{fig:max_age_multifigure_B}  we present the maximum age of pulsars in the detectable population as a function of the model parameters $\eta$ and $\epsilon_0$ for the case of Advanced Virgo and the Einstein Telescope. For detectable NSs in all models, for the current Advanced detectors, the maximum age is not larger than $1\,\rm{Myr}$ -- this is due to the fact that only the youngest NSs crossed the threshold of the detection. 

%Only in the case of very eccentric and long-lived pulsars in the top right corner of each graph the older stars up to $50\,\rm{Myr}$ can be seen. This is due to the fact that NSs have magnetic fields and their rotation periods slow down with time. Thus old NSs are slowly rotating and become invisible in GWs even if their ellipticity does not change. 
\begin{table}[]
    \centering
    \begin{tabular}{c|c|c|c|c|c}
        harmonic & $\rm{SNR}$ &  V1 & L1 & H1 & ET \\
    \hline
        $\nu$ & $11.4$ & 0 & 0.05 & 0.05 & 2.3 \\
        $2\nu$ & $11.4$ & 0 & 0.1 & 0.15 & 26.4 \\
    \hline
        $\nu$ &  $5$ & 0 & 0.1 & 0.05 & 7.15 \\
        $2\nu$ & $5$ & 0.1 & 0.7 & 0.5 & 82.0 \\
    \end{tabular}
    \caption{The expected number of detection for a model with parameters equal to: $\epsilon_{0}=10^{-5}$ and $\eta=10^{4}\,\rm{Myr}$. The numbers are normalised to an estimated number of Galactic NSs.}
    \label{tab:normalised_numbers}
\end{table}

\begin{figure*}
\centering
    \includegraphics[width=0.85\textwidth]{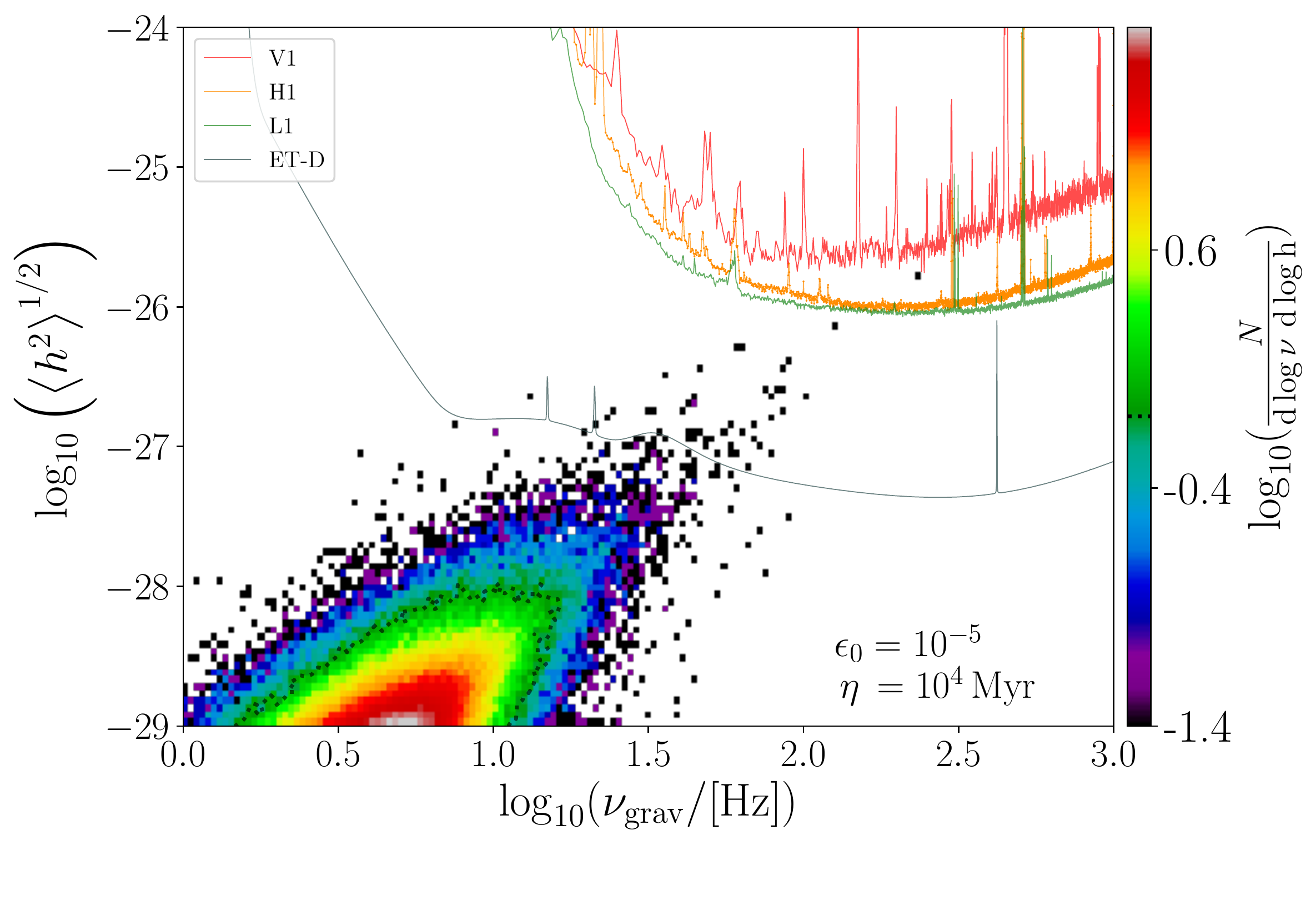}
    \vspace{-2.5em}
    \caption{The sensitivity curves for the Advanced LIGO (L1, H1), the Advanced Virgo (V1), and the Einstein Telescope in configuration D detectors for the signal's $\nu$ harmonic of the population of NSs. Model parameters equal to $\epsilon_{0}=10^{-5}$, and $\eta=10^{4}\rm{Myr}$. The sensitivity of the detectors is scaled by one year of integration time ($\times 1/\sqrt{t_{\rm{obs}}}$). The colour represents the density of NS population. The measured $h$ is estimated to for the ET.}
    \label{fig:MaxEpsMaxEta}
\end{figure*}

\begin{figure*}
\centering
    \includegraphics[width=0.85\textwidth]{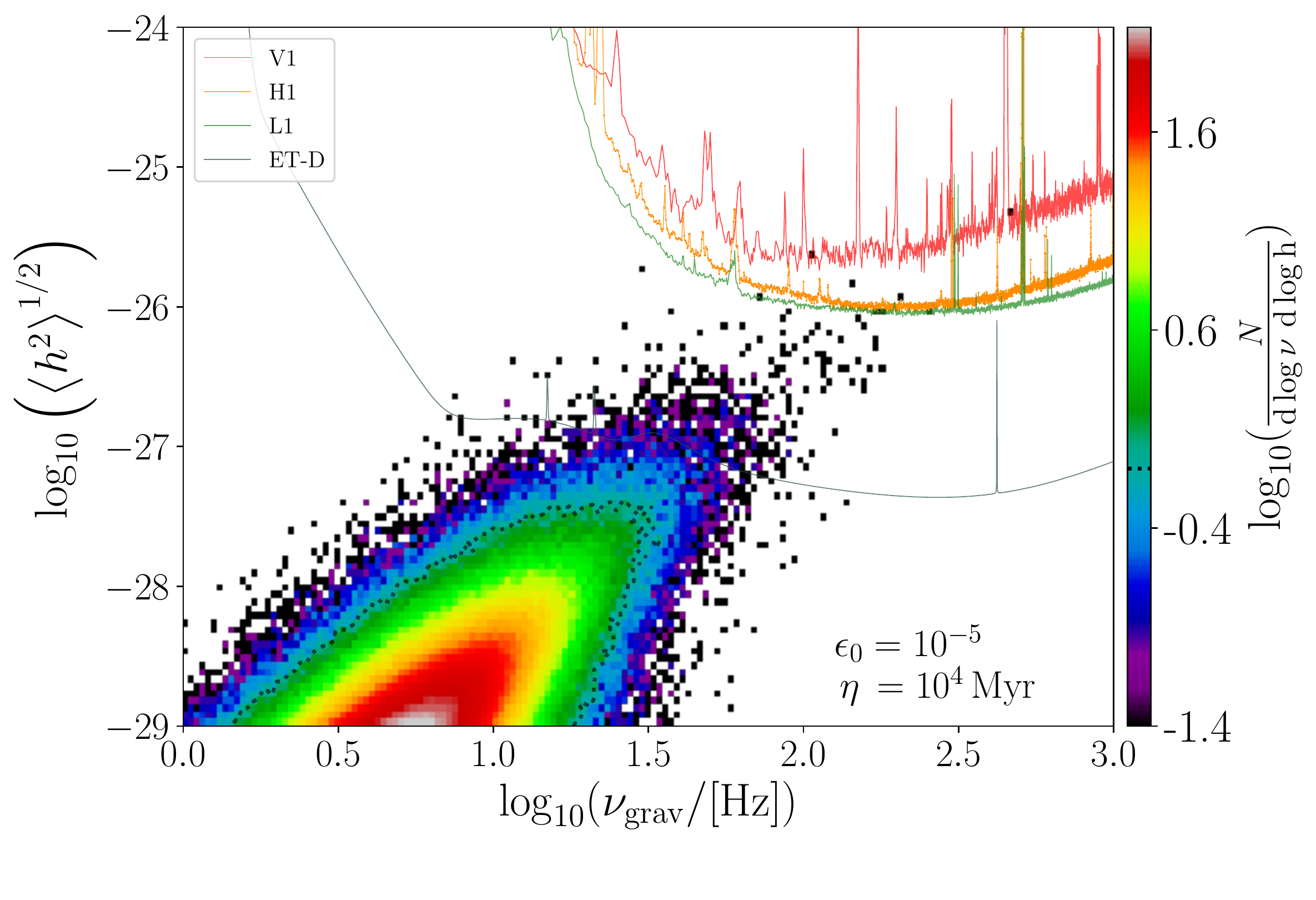}
    \vspace{-2.5em}
    \caption{The sensitivity curves for the Advanced LIGO (L1, H1), the Advanced Virgo (V1), and the Einstein Telescope in configuration D detectors for the signal's $2\nu$ harmonic of the population of NSs. Model parameters equal to $\epsilon_{0}=10^{-5}$, and $\eta=10^{4}\rm{Myr}$. The sensitivity of the detectors is scaled by one year of integration time ($\times 1/\sqrt{t_{\rm{obs}}}$). The colour represents the density of NS population. The measured $h$ is estimated to for the ET.}
    \label{fig:MaxEpsMaxEtaPhalf}
\end{figure*}

\begin{figure}
\centering

    \includegraphics[width=\columnwidth]{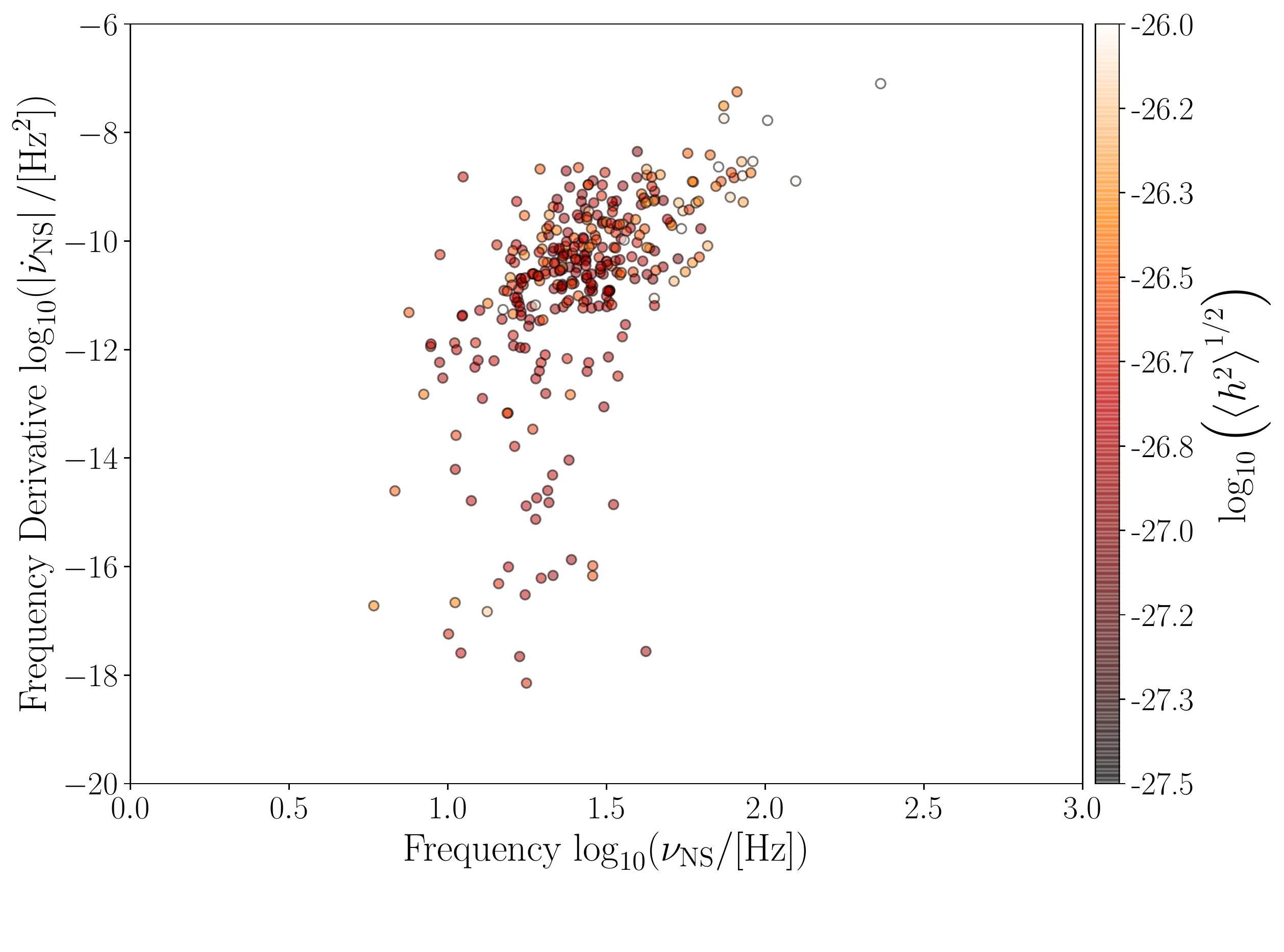}

    \caption{The frequency -- frequency derivative plane. The color represents the emitted GW at $2\nu$ frequency for the Einstein Telescope, configuration D, for one year of integration time. Model parameters equal to $\epsilon_{0}=10^{-5},\,\eta=10^{4}$. The presented population is $\times20$ larger than the expected Galactic population.}
    \label{fig:FFdotHED}
\end{figure}

\begin{figure}
\centering
    \includegraphics[width=\columnwidth]{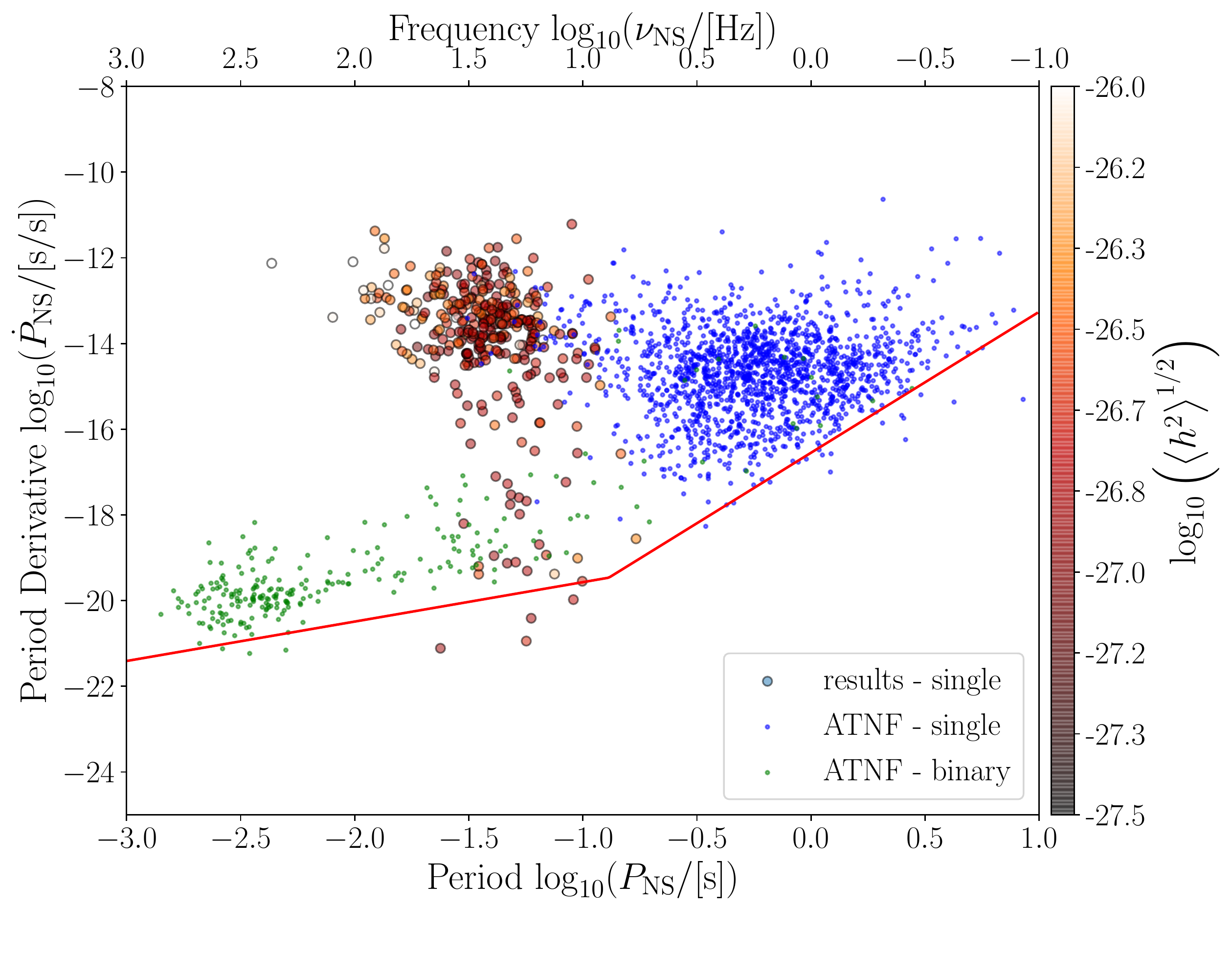}
    \caption{The period -- period derivative plane. The color represents the emitted GW at $2\nu$ frequency for the Einstein Telescope, configuration D, for one year of integration time. Model parameters equal to $\epsilon_{0}=10^{-5},\,\eta=10^{4}$. The presented population is $\times20$ higher than the expected Galactic population. The blue dots represent the population of observed, single pulsars. The green dots -- population of observed pulsars in binary systems (MSPs). Observed populations are from the Australia Telescope National Facility (ATNF) Pulsar Catalogue's \citep{2005AJ....129.1993M}. The red lines represent the {\em death lines} -- theoretical limit for an effective radio emission \citep{10.1093/mnras/267.3.513}.}
    \label{fig:PPdotED}
\end{figure}

\begin{figure}
\centering
    \begin{tabular}{c}
    \includegraphics[width=\columnwidth]{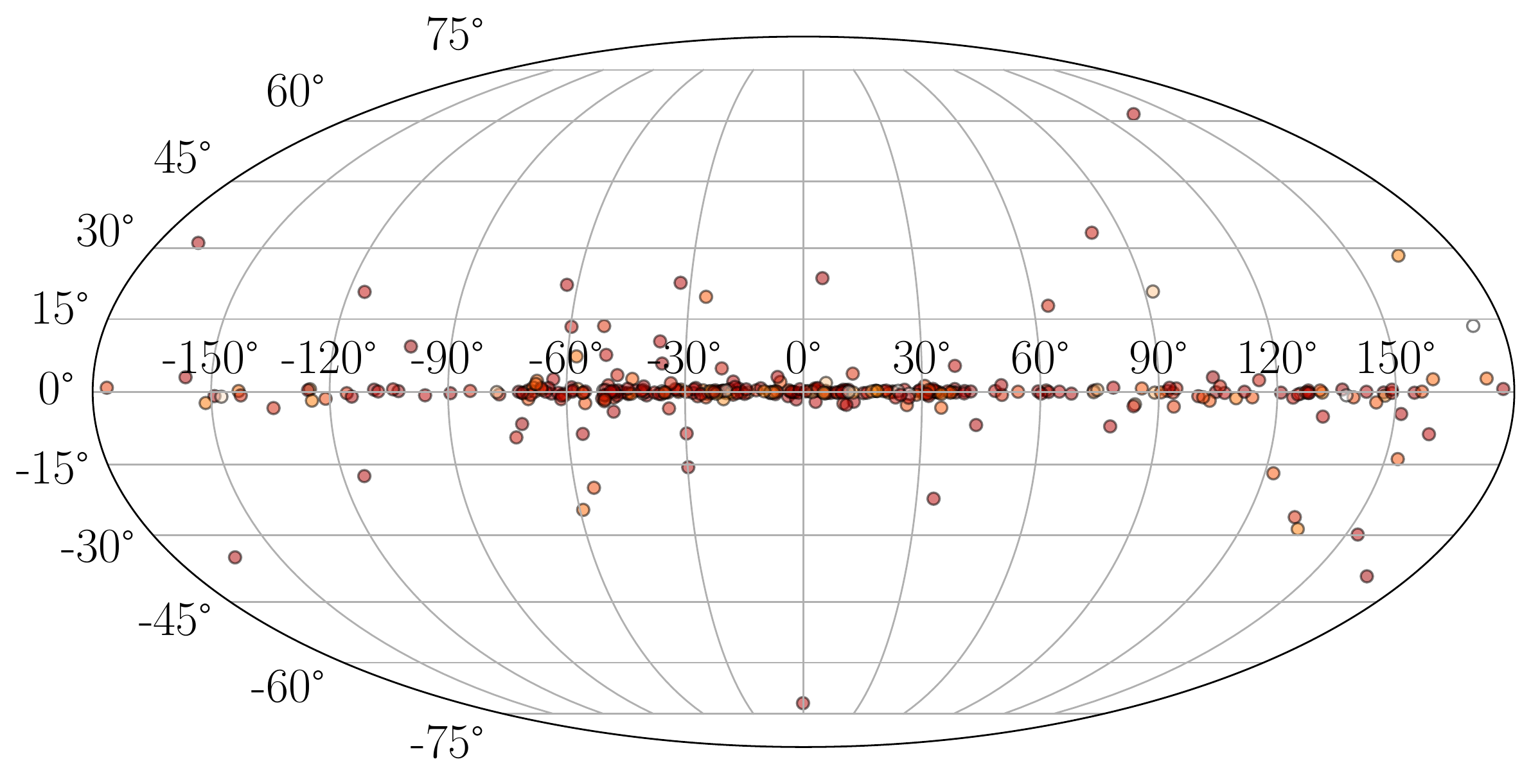} \\
    \includegraphics[width=0.9\columnwidth]{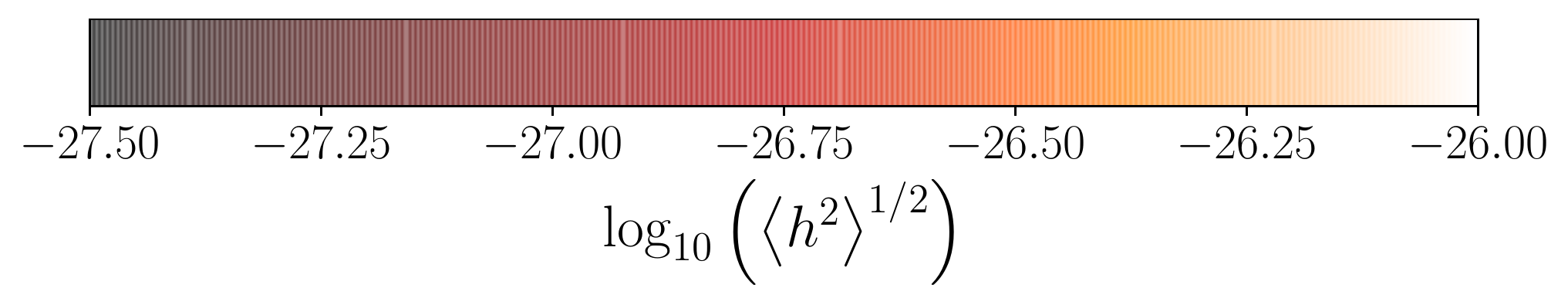}
    \end{tabular}
    \caption{Spatial distribution in the Galactic coordinates of visible population in the Einstein Telescope configuration D, at the $2\nu$ frequency, model parameters equal to $\epsilon_{0}=10^{-5},\,\eta=10^{4}$.}
    \label{fig:SpacialHED}
\end{figure}

Let us now concentrate on a single model with a $2.3$ and $26.4$ detections, for the $\nu$ and $2\nu$ harmonics respectively, in the Einstein Telescope described by $\epsilon_0 =10^{-5}$ and $\eta = 10^4\,\rm{Myr}$ (see Tab. \ref{tab:normalised_numbers}). 
In Figs. \ref{fig:MaxEpsMaxEta} and \ref{fig:MaxEpsMaxEtaPhalf}, we present the population of NSs on a diagram spanned by the GW frequency and the mean value of the GW amplitude. We also plot the detection threshold curves corresponding to one year integration and SNR threshold $11.4$ for the current detectors and for ET-D.
We note that the detectable populations will change by a factors ranging from 3 for the H1 detector to 7 for the L1 detector if the threshold is lowered to $\rm{SNR}=5$. In comparison, ET-D will increase its number of visible NSs only by a factor of 3. The discrepancy is due to the very poor statistics of the easiest to detect NSs, which leads to conclusion that a factor equal to 7 may be an overestimate. \\
%However, in the case of Einstein Telescope such change in the threshold can increase the detectable population even of a factor of $10$. 
The density of the population of NSs is shown as a color map. The observable NSs have frequencies in the range of approximately $10$ to $300\,\rm{Hz}$, with rare NSs going above $300\,\rm{Hz}$, when searching for the signal's $2\nu$ harmonic (see Fig. \ref{fig:FFdotHED}), and correspondingly lower for the signal's $\nu$ harmonic.
In Fig. \ref{fig:PPdotED} we compare the detectable, GW population (the color represents the amplitude of the signal's $2\nu$ harmonic) with the population of single pulsars in the Galaxy (based on the Australia Telescope National Facility Pulsar Catalogue's,  \citet{2005AJ....129.1993M}). The population of NSs that can be detected in ET-D resides mostly in the upper left corner of the period -- period derivative plane, which corresponds to the very young NSs. Since NSs are born in the Galactic disk, this leads to a spatially concentrated distribution (see Fig. \ref{fig:SpacialHED}).

 In Figure \ref{fig:FFdotHED} we present the population of pulsars detectable in ET in the space of variables used in gravitational wave searches: frequency and frequency derivative. The typical values of the frequency derivative $\dot{\nu}$ are in the range from $10^{-18}\,\rm{Hz}^2$ to $10^{-8}\,\rm{Hz}^2$. Thus most of the potentially detectable pulsars have frequency derivative that is not detectable. The frequencies of the detectable pulsars all lie below $300\,\rm{Hz}$, and the bulk of the detectable objects have frequencies in the range from $10\,\rm{Hz}$ to $60\,\rm{Hz}$. 

As a result, the optimal strategy to look for GW from isolated rotating NSs 
concentrate the surveys on the Galactic disk. In the case of of the Advanced detectors we expect detected pulsars only at $2\nu$ harmonic with frequencies in the range of $30$-$300\,\rm{Hz}$. In the case of the ET the bulk of detections will be in the frequency range from $20$ to $100\,\rm{Hz}$ for the $\nu$ harmonic and $10$ to $100\,\rm{Hz}$ for the $2\nu$ harmonic. 
The frequency derivatives of the brightest systems are above $10^{-8}\rm{Hz}^2$, while the typical values will be of frequency derivative in the population detectable by the ET is from $10^{-12}$ to $10^{-9}\rm{Hz^2}$. The largest value of the frequency derivative in the recent search \cite{2019PhRvD.100b4004A} was only $10^{-8}\rm{Hz}^2$.
Narrowing down the parameter space for the searches as suggested above  will increase the chances of detection.

\subsection{Limitation of the model -- Millisecond Pulsars}
\label{sec:MSPDisc}
If we inspect all detected pulsars (see green and blue populations on Fig. \ref{fig:PPdotED}) we note that the MSPs (green population) reside mostly in the frequency range from $200\rm{Hz}$ to $2\rm{kHz}$ for the more prominent harmonic of $f_{\rm{GW}}=2f_{\rm{NS}}$. Their addition could improve the detection prospects. However, answering question about quantitative improvement would require addressing the binary interactions (mention in \ref{sec:MSPModel} as well as more detailed model of asymmetry model that treats the initial and accretion induced inhomogeneity of the NS momentum. 

\section{Conclusions}
\begin{figure*}
\centering
\begin{tabular}{cc}
\begin{subfigure}{0.5\textwidth}\centering
    \includegraphics[width=1\columnwidth]{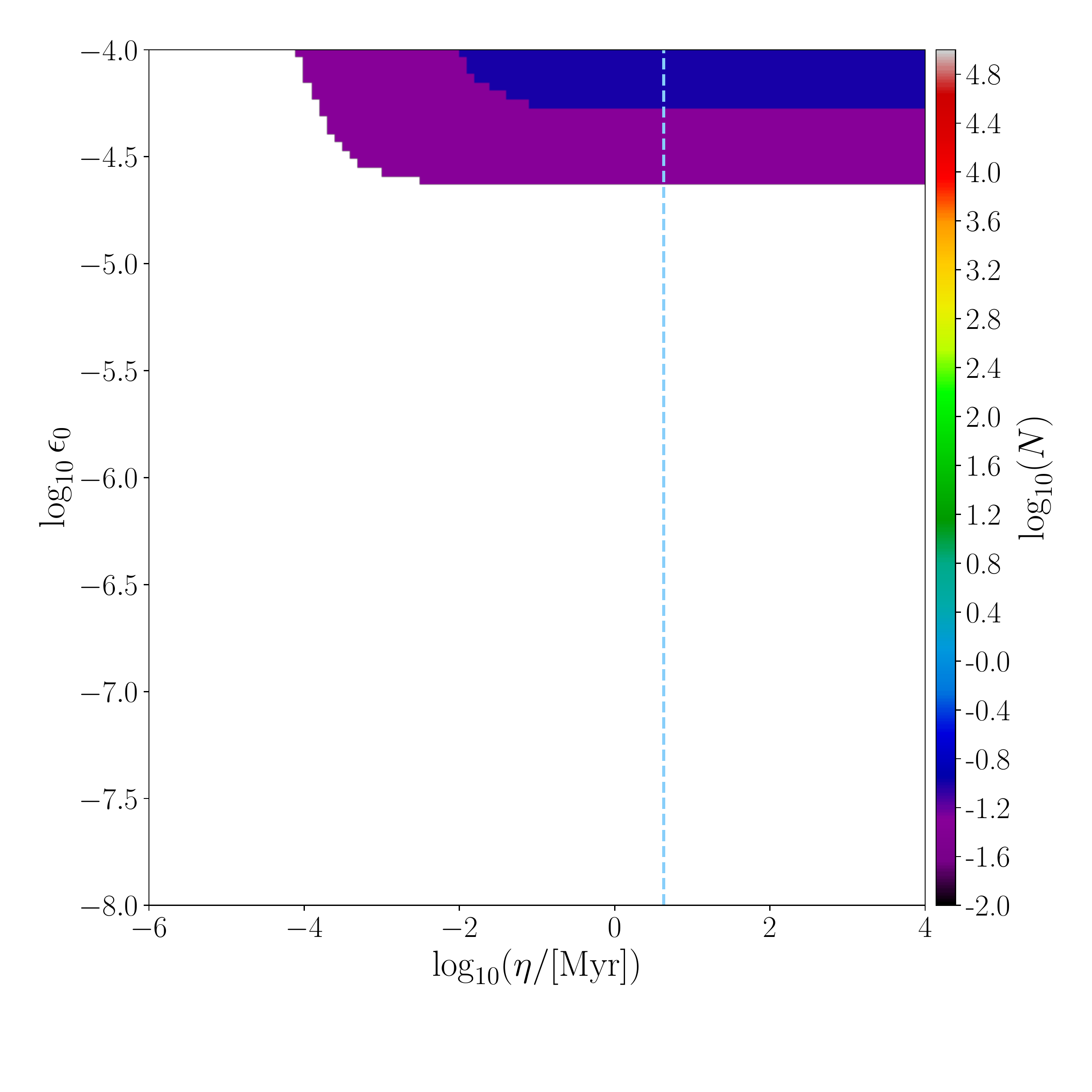}
    \vspace{-2em}
    \caption{$\nu$. V1 Advanced Virgo detector.}
    \label{fig:EtaEpsilonV1}
\end{subfigure}
&
\begin{subfigure}{0.5\textwidth}\centering
    \includegraphics[width=1\columnwidth]{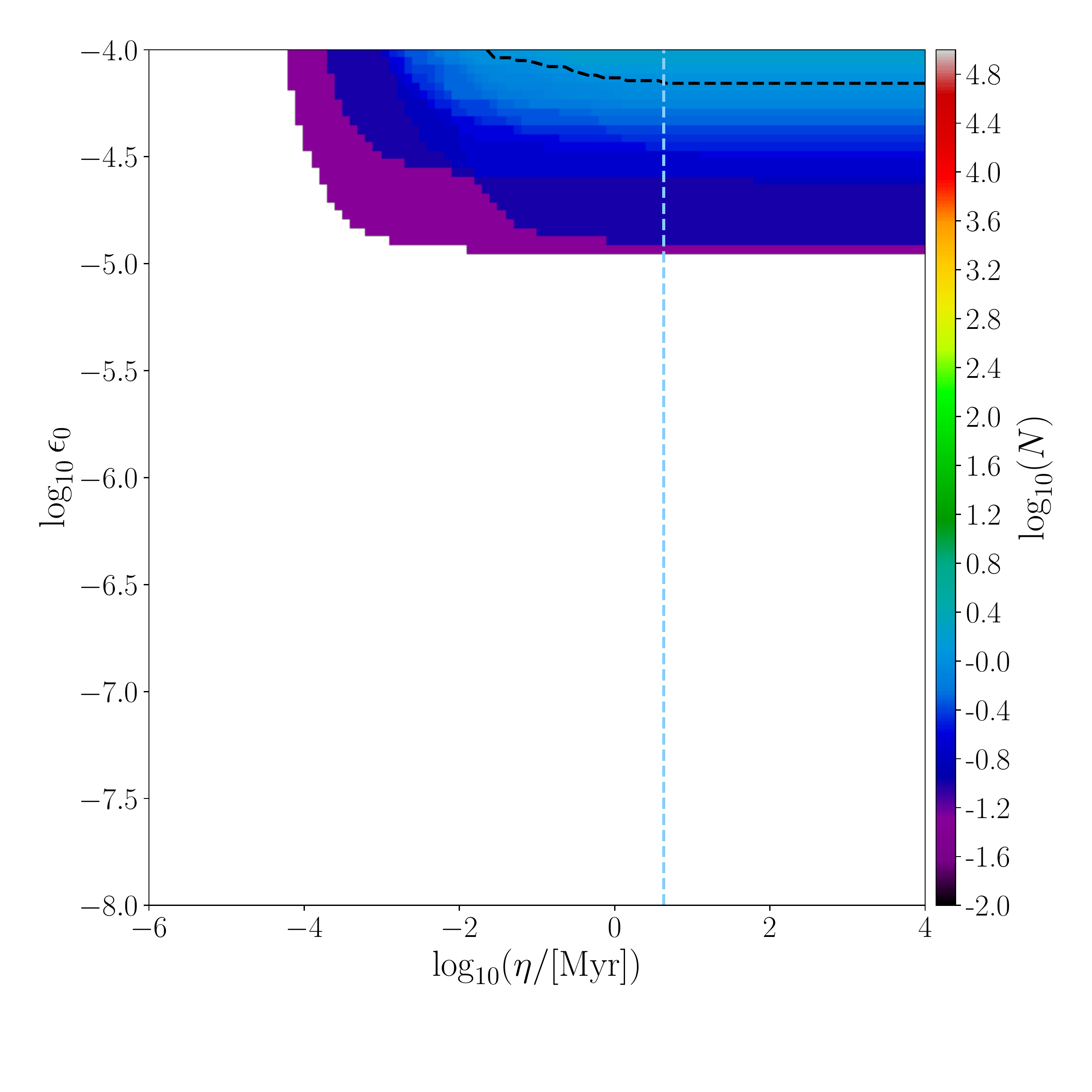}
    \vspace{-2em}
    \caption{$2\nu$. V1 Advanced Virgo detector.}
    \label{fig:EtaEpsilonV1_05}
\end{subfigure}
\\
\begin{subfigure}{0.5\textwidth}\centering
    \includegraphics[width=1\columnwidth]{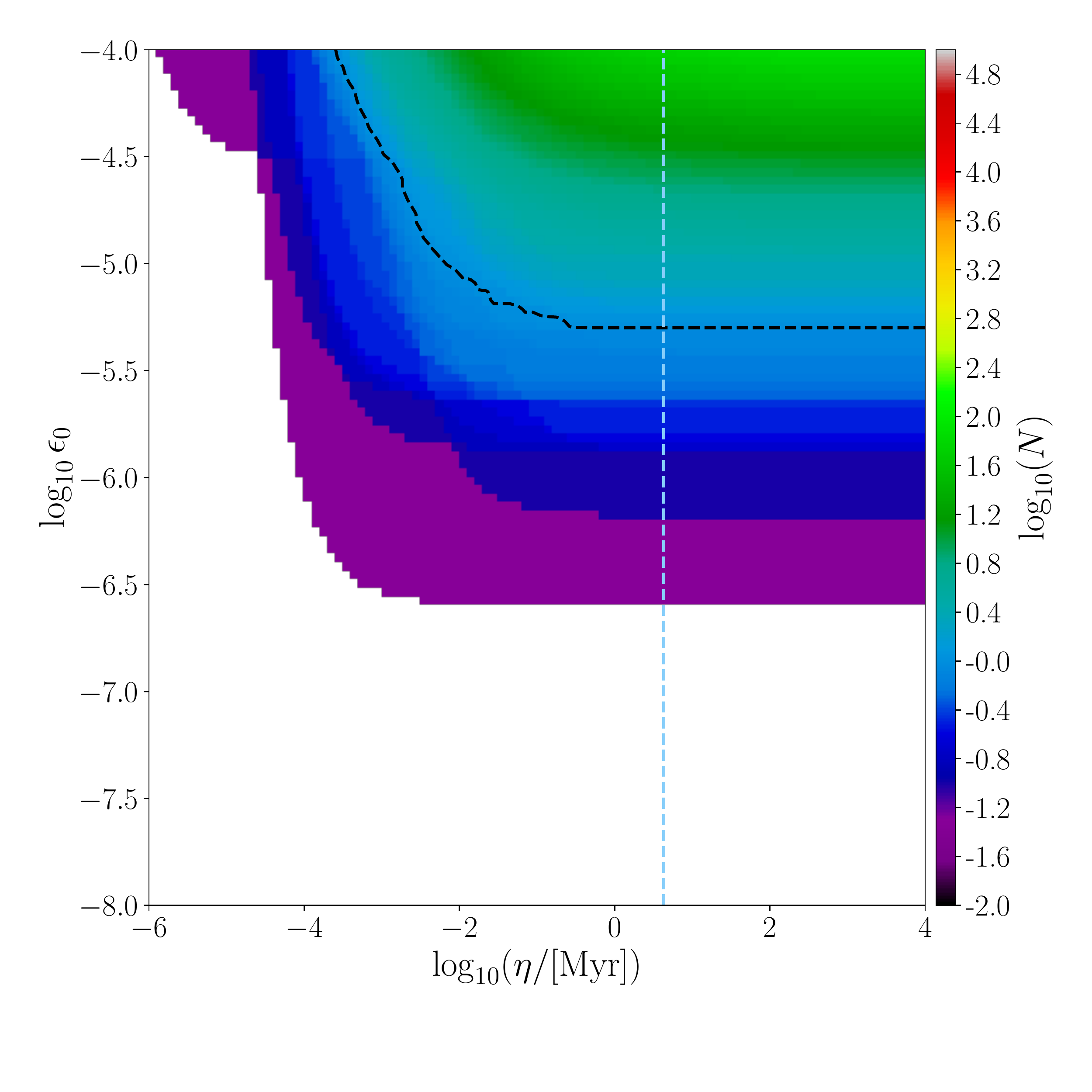}
    \vspace{-2em}
    \caption{$\nu$. ET-D Einstein Telescope detector, configuration D.}
    \label{fig:EtaEpsilonED}
\end{subfigure}
&
\begin{subfigure}{0.5\textwidth}\centering
    \includegraphics[width=1\columnwidth]{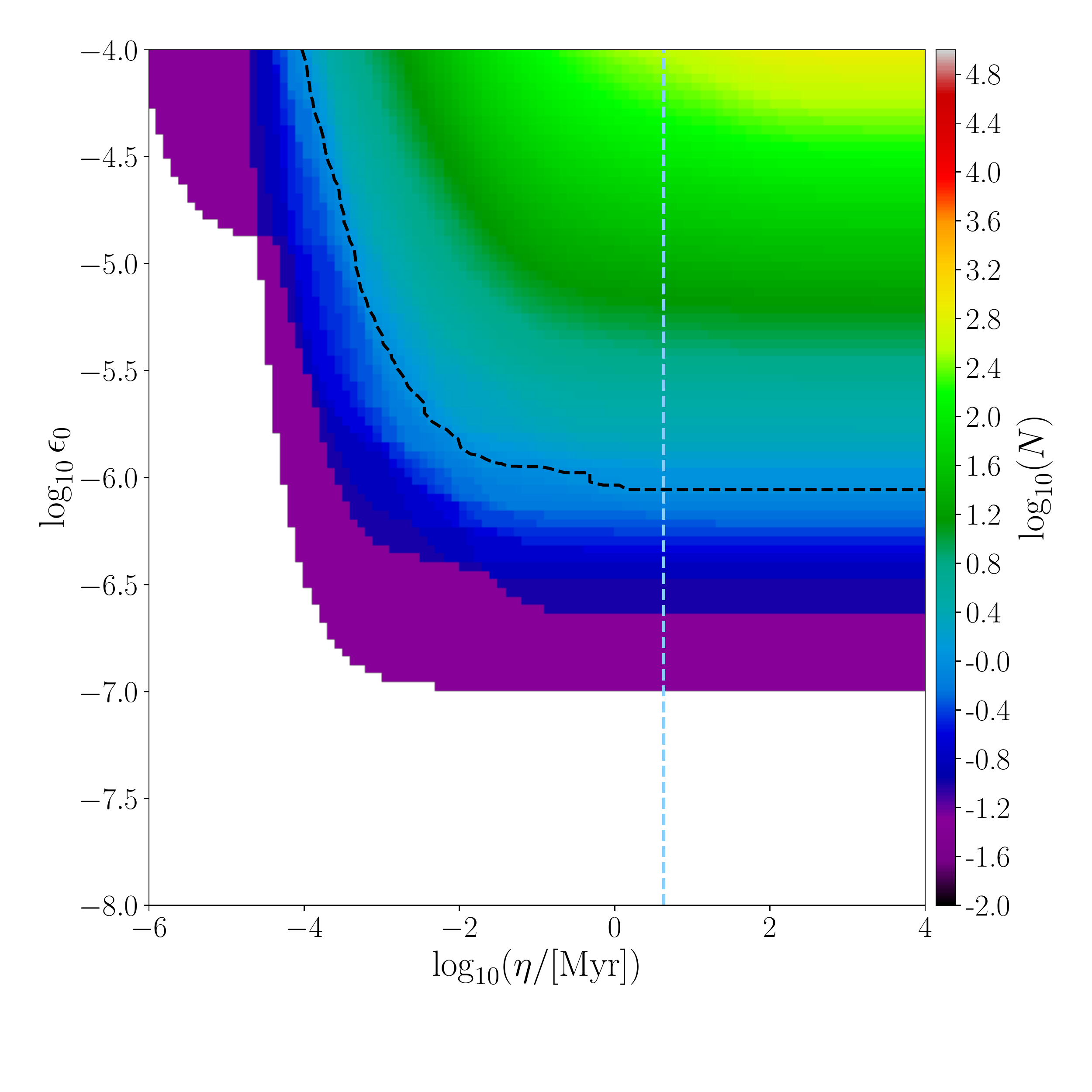}
    \vspace{-2em}
    \caption{$2\nu$. ET-D Einstein Telescope detector, configuration D.}
    \label{fig:EtaEpsilonED_05}
\end{subfigure}
\\
\end{tabular}
\caption{The number of observable pulsars in one year observations in the space of the parameters of the model $\eta-\epsilon_{0}$ for the Advanced Virgo, and Einstein Telescope detectors. The left column corresponds to the signal's $\nu$ harmonic, the right column the $2\nu$ harmonic. The black dashed line in each plot corresponds to the models where we expect to see one pulsar. All models below the black dashed line correspond to less than one expected detection. The blue, vertical, dashed line indicates models where $\eta$ is equal to the magnetic field decay $\Delta$ (see Tab. \ref{tab:params}).}
\label{fig:eta_epsilon_multifigure_B}
\end{figure*}

We presented our estimation of the detectability of the Galactic population of isolated NSs. 
With a high value of the initial ellipticity ($\epsilon_0 \simeq 10^{-5}$) and
no decay in the moment of inertia non-uniformity (decay scale $\eta = t_{\rm{hub}} 
\simeq 10^{4}\,\rm{Myr}$), the expected number of detected NSs in the Advanced Detectors is still less then $1$. 
Since the increase in signal-to-noise is proportional to square root of time 
(signal-to-noise $\sim\sqrt{t_{\rm{obs}}}$), we do not expect a drastic change in 
the estimates for the Advanced LIGO and Advanced Virgo detectors. The most limiting 
factor is the low frequency sensitivity (below $100\,\rm{Hz}$) of the detectors. As 
shown in Fig. \ref{fig:EtaEpsilonED}, we expect that future experiments such as 
the Einstein Telescope will clearly improve the prospect of continuous GW signal discovery.

We present the parameter space of the most likely discovery of solitary pulsars: the range frequencies, frequency derivatives and position in the sky. We suggest that narrowing down searches in this restricted parameter space may increase effective sensitivity and increase a chance of detection.

\begin{acknowledgements}
We would like to thank Brynmore Haskell and Graham Woan for comments and fruitful discussion. 

MC and TB are supported by the grant ``AstroCeNT: Particle Astrophysics Science and Technology Centre" (MAB/2018/7) carried out within the International Research Agendas programme of the Foundation for Polish Science (FNP) financed by the European Union under the European Regional Development Fund.
Part of this work was supported by Polish National Science Centre (NCN) grants no. 2016/22/E/ST9/00037 and 2017/26/M/ST9/00978. TB, MS, and NS acknowledges support of the TEAM/2016-3/19 grant from FNP. MS was partially supported by the NCN grant no. 2018/28/T/ST9/00458.
\end{acknowledgements}

\bibliographystyle{aa}
\bibliography{GWPopNS}

\begin{thebibliography}{59}
\expandafter\ifx\csname natexlab\endcsname\relax\def\natexlab#1{#1}\fi

\bibitem[{Aasi {et~al.}(2015)Aasi, Abbott, Abbott, Abbott, Abernathy, Ackley,
  Adams, Adams, Addesso, Adhikari, Adya, Affeldt, Aggarwal, Aguiar, Ain, Ajith,
  Alemic, Allen, Amariutei, Anderson, Anderson, Arai, Araya, Arceneaux, Areeda,
  Ashton, Ast, Aston, Aufmuth, Aulbert, Aylott, Babak, Baker, Ballmer,
  Barayoga, Barbet, Barclay, Barish, Barker, Barr, Barsotti, Bartlett, Barton,
  Bartos, Bassiri, Batch, Baune, Behnke, Bell, Bell, Benacquista, Bergman,
  Bergmann, Berry, Betzwieser, Bhagwat, Bhandare, Bilenko, Billingsley, Birch,
  Biscans, Biwer, Blackburn, Blackburn, Blair, Blair, Bock, Bodiya, Bojtos,
  Bond, {et~al.}}]{AdvLIGO2015}
Aasi, J., Abbott, B.~P., Abbott, R., {et~al.} 2015, Classical and Quantum
  Gravity, 32, 074001

\bibitem[{Aasi {et~al.}(2016)Aasi, Abbott, Abbott, Abbott, Abernathy, Acernese,
  Ackley, Adams, Adams, Addesso, Adhikari, Adya, Affeldt, Agathos, Agatsuma,
  Aggarwal, Aguiar, Ain, Ajith, Allen, Allocca, Amariutei, Andersen, Anderson,
  Anderson, Arai, Araya, Arceneaux, Areeda, Arnaud, Ashton, Aston, Astone,
  Aufmuth, Aulbert, Babak, Baker, Baldaccini, Ballardin, Ballmer, Barayoga,
  Barclay, Barish, Barker, Barone, Barr, Barsotti, Barsuglia, Bartlett, Barton,
  Bartos, Bassiri, Basti, Batch, Baune, Bavigadda, Behnke, Bejger, Belczynski,
  Bell, Berger, Bergman, Bergmann, Berry, Bersanetti, Bertolini, Betzwieser,
  Bhagwat, Bhandare, Bilenko, Billingsley, Birch, Birney, Biscans, Bitossi,
  Biwer, Bizouard, Blackburn, Blair, Blair, Bloemen, Bock, Bodiya, Boer,
  Bogaert, Bojtos, Bond, Bondu, Bonnand, Bork, Born, Boschi, Bose, Bradaschia,
  Brady, Braginsky, Branchesi, Branco, Brau, Briant, Brillet, Brinkmann,
  Brisson, Brockill, Brooks, Brown, Brown, Brown, Brown, Buchanan, Buikema,
  Bulik, Bulten, Buonanno, Buskulic, Buy, Byer, Cadonati, Cagnoli,
  Calder\'on~Bustillo, Calloni, Camp, Cannon, Cao, Capano, Capocasa,
  Carbognani, Caride, Casanueva~Diaz, Casentini, Caudill, Cavagli\`a, Cavalier,
  Cavalieri, Celerier, Cella, Cepeda, Cerboni~Baiardi, Cerretani, Cesarini,
  Chakraborty, Chalermsongsak, Chamberlin, Chao, Charlton, Chassande-Mottin,
  Chen, Chen, Cheng, Chincarini, Chiummo, Cho, Cho, Chow, Christensen, Chu,
  Chua, Chung, Ciani, Clara, Clark, Cleva, Coccia, Cohadon, Colla, Collette,
  Colombini, Constancio, Conte, Conti, Cook, Corbitt, Cornish, Corsi, Costa,
  Coughlin, Coughlin, Coulon, Countryman, Couvares, Coward, Cowart, Coyne,
  Coyne, Craig, Creighton, Cripe, Crowder, Cumming, Cunningham, Cuoco, Canton,
  Damjanic, Danilishin, D'Antonio, Danzmann, Darman, Dattilo, Dave, Daveloza,
  Davier, Davies, Daw, Day, DeBra, Debreczeni, Degallaix, De~Laurentis,
  Del\'eglise, Del~Pozzo, Denker, Dent, Dereli, Dergachev, De~Rosa, DeRosa,
  DeSalvo, Dhurandhar, D\'{\i}az, Di~Fiore, Di~Giovanni, Di~Lieto, Di~Palma,
  Di~Virgilio, Dojcinoski, Dolique, Dominguez, Donovan, Dooley, Doravari,
  Douglas, Downes, Drago, Drever, Driggers, Du, Ducrot, Dwyer, Edo, Edwards,
  Edwards, Effler, Eggenstein, Ehrens, Eichholz, Eikenberry, Essick, Etzel,
  Evans, Evans, Everett, Factourovich, Fafone, Fairhurst, Fang, Farinon, Farr,
  Farr, Favata, Fays, Fehrmann, Fejer, Feldbaum, Ferrante, Ferreira, Ferrini,
  Fidecaro, Fiori, Fisher, Flaminio, Fournier, Franco, Frasca, Frasconi, Frede,
  Frei, Freise, Frey, Fricke, Fritschel, Frolov, Fulda, Fyffe, Gabbard, Gair,
  Gammaitoni, Gaonkar, Garufi, Gatto, Gehrels, Gemme, Gendre, Genin, Gennai,
  Gergely, Germain, Ghosh, Ghosh, Giaime, Giardina, Giazotto, Gleason, Goetz,
  Goetz, Gondan, Gonz\'alez, Gonzalez, Gopakumar, Gordon, Gorodetsky, Gossan,
  Gosselin, Go\ss{}ler, Gouaty, Graef, Graff, Granata, Grant, Gras, Gray,
  Greco, Groot, Grote, Grover, Grunewald, Guidi, Guido, Guo, Gupta, Gupta,
  Gushwa, Gustafson, Gustafson, Hacker, Hall, Hall, Hammer, Hammond, Haney,
  Hanke, Hanks, Hanna, Hannam, Hanson, Hardwick, Harms, Harry, Harry, Hart,
  Hartman, Haster, Haughian, Heidmann, Heintze, Heitmann, Hello, Hemming,
  Hendry, Heng, Hennig, Heptonstall, Heurs, Hild, Hoak, Hodge,
  Hoelscher-Obermaier, Hofman, Hollitt, Holt, Hopkins, Hosken, Hough, Houston,
  Howell, Hu, Huang, Huerta, Huet, Hughey, Husa, Huttner, Huynh, Huynh-Dinh,
  Idrisy, Indik, Ingram, Inta, Islas, Isler, Isogai, Iyer, Izumi, Jacobson,
  Jang, Jaranowski, Jawahar, Ji, Jim\'enez-Forteza, Johnson, Jones, Jones,
  Jonker, Ju, Haris, Kalogera, Kandhasamy, Kang, Kanner, Karki, Karlen,
  Kasprzack, Katsavounidis, Katzman, Kaufer, Kaur, Kawabe, Kawazoe,
  K\'ef\'elian, Kehl, Keitel, Kelley, Kells, Kerrigan, Key, Khalili, Khan,
  Khazanov, Kijbunchoo, Kim, Kim, Kim, Kim, Kim, King, King, Kinzel, Kissel,
  Klimenko, Kline, Koehlenbeck, Kokeyama, Koley, Kondrashov, Korobko, Korth,
  Kowalska, Kozak, Kringel, Krishnan, Kr\'olak, Krueger, Kuehn, Kumar, Kumar,
  Kuo, Kutynia, Lackey, Landry, Lantz, Lasky, Lazzarini, Lazzaro, Leaci,
  Leavey, Lebigot, Lee, Lee, Lee, Lee, Lee, Leonardi, Leong, Leroy, Letendre,
  Levin, Levine, Lewis, Li, Libson, Lin, Littenberg, Lockerbie, Lockett,
  Lodhia, Logue, Lombardi, Lorenzini, Loriette, Lormand, Losurdo, Lough,
  Lubinski, L\"uck, Lundgren, Luo, Lynch, Ma, Macarthur, Macdonald, MacDonald,
  Machenschalk, MacInnis, Macleod, Madden-Fong, Maga\~na Sandoval, Magee,
  Mageswaran, Majorana, Maksimovic, Malvezzi, Man, Mandel, Mandic, Mangano,
  Mangini, Mansell, Manske, Mantovani, Marchesoni, Marion, M\'arka, M\'arka,
  Markosyan, Maros, Martelli, Martellini, Martin, Martin, Martynov, Marx,
  Mason, Masserot, Massinger, Mastrogiovanni, Matichard, Matone, Mavalvala,
  Mazumder, Mazzolo, McCarthy, McClelland, McCormick, McGuire, McIntyre,
  McIver, McWilliams, Meacher, Meadors, Mehmet, Meidam, Meinders, Melatos,
  Mendell, Mercer, Merzougui, Meshkov, Messenger, Messick, Meyers, Mezzani,
  Miao, Michel, Middleton, Mikhailov, Milano, Miller, Millhouse, Minenkov,
  Ming, Mirshekari, Mishra, Mitra, Mitrofanov, Mitselmakher, Mittleman, Moe,
  Moggi, Mohan, Mohapatra, Montani, Moore, Moraru, Moreno, Morriss, Mossavi,
  Mours, Mow-Lowry, Mueller, Mueller, Mukherjee, Mukherjee, Mullavey, Munch,
  Murphy, Murray, Mytidis, Nagy, Nardecchia, Naticchioni, Nayak, Necula,
  Nedkova, Nelemans, Neri, Newton, Nguyen, Nielsen, Nitz, Nocera, Nolting,
  Normandin, Nuttall, Ochsner, O'Dell, Oelker, Ogin, Oh, Oh, Ohme, Okounkova,
  Oppermann, Oram, O'Reilly, Ortega, O'Shaughnessy, Ott, Ottaway, Ottens,
  Overmier, Owen, Padilla, Pai, Pai, Palamos, Palashov, Palomba, Pal-Singh,
  Pan, Pan, Pankow, Pannarale, Pant, Paoletti, Papa, Paris, Pasqualetti,
  Passaquieti, Passuello, Patrick, Pedraza, Pekowsky, Pele, Penn, Perreca,
  Phelps, Piccinni, Pichot, Pickenpack, Piergiovanni, Pierro, Pillant, Pinard,
  Pinto, Pitkin, Poeld, Poggiani, Post, Powell, Prasad, Predoi, Premachandra,
  Prestegard, Price, Prijatelj, Principe, Privitera, Prix, Prodi, Prokhorov,
  Puncken, Punturo, Puppo, P\"urrer, Qin, Quetschke, Quintero, Quitzow-James,
  Raab, Rabeling, R\'acz, Radkins, Raffai, Raja, Rakhmanov, Rapagnani, Raymond,
  Razzano, Re, Reed, Regimbau, Rei, Reid, Reitze, Ricci, Riles, Robertson,
  Robie, Robinet, Rocchi, Rodger, Rolland, Rollins, Roma, Romano, Romano,
  Romanov, Romie, Rosi\ifmmode~\acute{n}\else \'{n}\fi{}ska, Rowan, R\"udiger,
  Ruggi, Ryan, Sachdev, Sadecki, Sadeghian, Saleem, Salemi, Sammut, Sanchez,
  Sandberg, Sanders, Santiago-Prieto, Sassolas, Sathyaprakash, Saulson, Savage,
  Sawadsky, Schale, Schilling, Schmidt, Schnabel, Schofield, Sch\"onbeck,
  Schreiber, Schuette, Schutz, Scott, Scott, Sellers, Sentenac, Sequino,
  Sergeev, Serna, Sevigny, Shaddock, Shaffery, Shah, Shahriar, Shaltev, Shao,
  Shapiro, Shawhan, Shoemaker, Sidery, Siellez, Siemens, Sigg, Silva, Simakov,
  Singer, Singer, Singh, Sintes, Slagmolen, Smith, Smith, Smith, Son, Sorazu,
  Souradeep, Srivastava, Staley, Steinke, Steinlechner, Steinlechner,
  Steinmeyer, Stephens, Steplewski, Stevenson, Stone, Strain, Straniero,
  Strauss, Strigin, Sturani, Stuver, Summerscales, Sun, Sutton, Swinkels,
  Szczepanczyk, Tacca, Talukder, Tanner, T\'apai, Tarabrin, Taracchini, Taylor,
  Theeg, Thirugnanasambandam, Thomas, Thomas, Thorne, Thorne, Thrane, Tiwari,
  Tiwari, Tokmakov, Tomlinson, Tonelli, Torres, Torrie, Travasso, Traylor,
  Trifir\`o, Tringali, Tse, Turconi, Ugolini, Unnikrishnan, Urban, Usman,
  Vahlbruch, Vajente, Valdes, Vallisneri, van Bakel, van Beuzekom, van~den
  Brand, van~den Broeck, van~der Schaaf, van~der Sluys, van Heijningen, van
  Veggel, Vardaro, Vass, Vas\'uth, Vaulin, Vecchio, Vedovato, Veitch, Veitch,
  Venkateswara, Verkindt, Vetrano, Vicer\'e, Vinet, Vitale, Vo, Vocca, Vorvick,
  Vousden, Vyatchanin, Wade, Wade, Wade, Walker, Wallace, Walsh, Wang, Wang,
  Wang, Wang, Ward, Warner, Was, Weaver, Wei, Weinert, Weinstein, Weiss,
  Welborn, Wen, We\ss{}els, Westphal, Wette, Whelan, Whitcomb, White, Whiting,
  Williams, Williams, Williams, Williamson, Willis, Willke, Wimmer, Winkler,
  Wipf, Wittel, Woan, Worden, Yablon, Yakushin, Yam, Yamamoto, Yancey, Yvert,
  Zadro\ifmmode~\dot{z}\else \.{z}\fi{}ny, Zangrando, Zanolin, Zendri, Zhang,
  Zhang, Zhang, Zhang, Zhao, Zhou, Zhu, Zucker, Zuraw, \&
  Zweizig}]{PhysRevD.93.042007}
Aasi, J., Abbott, B.~P., Abbott, R., {et~al.} 2016, Phys. Rev. D, 93, 042007

\bibitem[{Abbott {et~al.}(2009)Abbott, Abbott, Adhikari, Ajith, Allen, Allen,
  Amin, Anderson, Anderson, Anderson, Arain, Araya, Armandula, Armor, Aso,
  Aston, Aufmuth, {et~al.}}]{2e829abe62dc4158b2b0e895b23c2c1d}
Abbott, B., Abbott, R., Adhikari, R., {et~al.} 2009, Physical Review D -
  Particles, Fields, Gravitation and Cosmology, 79

\bibitem[{{Abbott} {et~al.}(2018{\natexlab{a}}){Abbott}, {Abbott}, {Abbott},
  {Abernathy}, {Acernese}, {Ackley}, {Adams}, {Adams}, {Addesso}, {Adhikari},
  {Adya}, {Affeldt}, {Agathos}, {Agatsuma}, {Aggarwal}, {Aguiar}, {Aiello},
  {Ain}, {Ajith}, {Akutsu}, {Allen}, {Allocca}, {Altin}, {Ananyeva},
  {Anderson}, {Anderson}, {Ando}, {Appert}, {Arai}, {Araya}, {Araya}, {Areeda},
  {Arnaud}, {Arun}, {Asada}, {Ascenzi}, {Ashton}, {Aso}, {Ast}, {Aston},
  {Astone}, {Atsuta}, {Aufmuth}, {Aulbert}, {Avila-Alvarez}, {Awai},
  {et~al.}}]{2018LRR....21....3A}
{Abbott}, B.~P., {Abbott}, R., {Abbott}, T.~D., {et~al.} 2018{\natexlab{a}},
  Living Reviews in Relativity, 21, 3

\bibitem[{{Abbott} {et~al.}(2018{\natexlab{b}}){Abbott}, {Abbott}, {Abbott},
  {Abernathy}, {Acernese}, {Ackley}, {Adams}, {Adams}, {Addesso}, {Adhikari},
  {Adya}, {Affeldt}, {Agathos}, {Agatsuma}, {Aggarwal}, {Aguiar}, {Aiello},
  {Ain}, {Ajith}, {Allen}, {Allocca}, {Altin}, {Ananyeva}, {Anderson},
  {Anderson}, {Appert}, {et~al.}}]{2018CQGra..35f5009A}
{Abbott}, B.~P., {Abbott}, R., {Abbott}, T.~D., {et~al.} 2018{\natexlab{b}},
  Classical and Quantum Gravity, 35, 065009

\bibitem[{Abbott {et~al.}(2016)Abbott, Abbott, Abbott, Abernathy, Acernese,
  Ackley, Adams, Adams, Addesso, Adhikari, Adya, Affeldt, Agathos, Agatsuma,
  Aggarwal, Aguiar, Aiello, Ain, Ajith, Allen, Allocca, Altin, Anderson,
  Anderson, Arai, Arain, Araya, Arceneaux, Areeda, Arnaud, Arun, Ascenzi,
  Ashton, Ast, Aston, Astone, Aufmuth, Aulbert, Babak, Bacon, Bader, Baker,
  Baldaccini, Ballardin, Ballmer, Barayoga, Barclay, Barish, Barker, Barone,
  Barr, Barsotti, Barsuglia, Barta, Bartlett, Barton, Bartos, Bassiri, Basti,
  Batch, Baune, Bavigadda, Bazzan, Behnke, Bejger, Belczynski, Bell, Bell,
  Berger, Bergman, Bergmann, Berry, Bersanetti, Bertolini, Betzwieser, Bhagwat,
  Bhandare, Bilenko, Billingsley, Birch, Birney, Birnholtz, Biscans, Bisht,
  Bitossi, Biwer, Bizouard, Blackburn, Blair, Blair, Blair, Bloemen, Bock,
  Bodiya, Boer, Bogaert, Bogan, Bohe, Bojtos, Bond, Bondu, Bonnand, Boom, Bork,
  Boschi, Bose, Bouffanais, Bozzi, Bradaschia, Brady, Braginsky, Branchesi,
  Brau, Briant, Brillet, Brinkmann, Brisson, Brockill, Brooks, Brown, Brown,
  Brown, Buchanan, Buikema, Bulik, Bulten, Buonanno, Buskulic, Buy, Byer,
  Cabero, Cadonati, Cagnoli, Cahillane, Bustillo, Callister, Calloni, Camp,
  Cannon, Cao, Capano, Capocasa, Carbognani, Caride, Diaz, Casentini, Caudill,
  Cavagli\`a, Cavalier, Cavalieri, Cella, Cepeda, Baiardi, Cerretani, Cesarini,
  Chakraborty, Chalermsongsak, Chamberlin, Chan, Chao, Charlton,
  Chassande-Mottin, Chen, Chen, Cheng, Chincarini, Chiummo, Cho, Cho, Chow,
  Christensen, Chu, Chua, Chung, Ciani, Clara, Clark, Cleva, Coccia, Cohadon,
  Colla, Collette, Cominsky, Constancio, Conte, Conti, Cook, Corbitt, Cornish,
  Corsi, Cortese, Costa, Coughlin, Coughlin, Coulon, Countryman, Couvares,
  Cowan, Coward, Cowart, Coyne, Coyne, Craig, Creighton, Creighton, Cripe,
  Crowder, Cruise, Cumming, Cunningham, Cuoco, Canton, Danilishin, D'Antonio,
  Danzmann, Darman, Da~Silva~Costa, Dattilo, Dave, Daveloza, Davier, Davies,
  Daw, Day, De, DeBra, Debreczeni, Degallaix, De~Laurentis, Del\'eglise,
  Del~Pozzo, Denker, Dent, Dereli, Dergachev, DeRosa, De~Rosa, DeSalvo,
  Dhurandhar, D\'{\i}az, Di~Fiore, Di~Giovanni, Di~Lieto, Di~Pace, Di~Palma,
  Di~Virgilio, Dojcinoski, Dolique, Donovan, Dooley, Doravari, Douglas, Downes,
  Drago, Drever, Driggers, Du, Ducrot, Dwyer, Edo, Edwards, Effler, Eggenstein,
  Ehrens, Eichholz, Eikenberry, Engels, Essick, Etzel, Evans, Evans, Everett,
  Factourovich, Fafone, Fair, Fairhurst, Fan, Fang, Farinon, Farr, Farr,
  Favata, Fays, Fehrmann, Fejer, Feldbaum, Ferrante, Ferreira, Ferrini,
  Fidecaro, Finn, Fiori, Fiorucci, Fisher, Flaminio, Fletcher, Fong, Fournier,
  Franco, Frasca, Frasconi, Frede, Frei, Freise, Frey, Frey, Fricke, Fritschel,
  Frolov, Fulda, Fyffe, Gabbard, Gair, Gammaitoni, Gaonkar, Garufi, Gatto,
  Gaur, Gehrels, Gemme, Gendre, Genin, Gennai, George, Gergely, Germain, Ghosh,
  Ghosh, Ghosh, Giaime, Giardina, Giazotto, Gill, Glaefke, Gleason, Goetz,
  Goetz, Gondan, Gonz\'alez, Castro, Gopakumar, Gordon, Gorodetsky, Gossan,
  Gosselin, Gouaty, Graef, Graff, Granata, Grant, Gras, Gray, Greco, Green,
  Greenhalgh, Groot, Grote, Grunewald, Guidi, Guo, Gupta, Gupta, Gushwa,
  Gustafson, Gustafson, Hacker, Hall, Hall, Hammond, Haney, Hanke, Hanks,
  Hanna, Hannam, Hanson, Hardwick, Harms, Harry, Harry, Hart, Hartman, Haster,
  Haughian, Healy, Heefner, Heidmann, Heintze, Heinzel, Heitmann, Hello,
  Hemming, Hendry, Heng, Hennig, Heptonstall, Heurs, Hild, Hoak, Hodge, Hofman,
  Hollitt, Holt, Holz, Hopkins, Hosken, Hough, Houston, Howell, Hu, Huang,
  Huerta, Huet, Hughey, Husa, Huttner, Huynh-Dinh, Idrisy, Indik, Ingram, Inta,
  Isa, Isac, Isi, Islas, Isogai, Iyer, Izumi, Jacobson, Jacqmin, Jang, Jani,
  Jaranowski, Jawahar, Jim\'enez-Forteza, Johnson, Johnson-McDaniel, Jones,
  Jones, Jonker, Ju, Haris, Kalaghatgi, Kalogera, Kandhasamy, Kang, Kanner,
  Karki, Kasprzack, Katsavounidis, Katzman, Kaufer, Kaur, Kawabe, Kawazoe,
  K\'ef\'elian, Kehl, Keitel, Kelley, Kells, Kennedy, Keppel, Key,
  Khalaidovski, Khalili, Khan, Khan, Khan, Khazanov, Kijbunchoo, Kim, Kim, Kim,
  Kim, Kim, Kim, King, King, Kinzel, Kissel, Kleybolte, Klimenko, Koehlenbeck,
  Kokeyama, Koley, Kondrashov, Kontos, Koranda, Korobko, Korth, Kowalska,
  Kozak, Kringel, Krishnan, Kr\'olak, Krueger, Kuehn, Kumar, Kumar, Kuo,
  Kutynia, Kwee, Lackey, Landry, Lange, Lantz, Lasky, Lazzarini, Lazzaro,
  Leaci, Leavey, Lebigot, Lee, Lee, Lee, Lee, Lenon, Leonardi, Leong, Leroy,
  Letendre, Levin, Levine, Li, Libson, Littenberg, Lockerbie, Logue, Lombardi,
  London, Lord, Lorenzini, Loriette, Lormand, Losurdo, Lough, Lousto, Lovelace,
  L\"uck, Lundgren, Luo, Lynch, Ma, MacDonald, Machenschalk, MacInnis, Macleod,
  Maga\~na Sandoval, Magee, Mageswaran, Majorana, Maksimovic, Malvezzi, Man,
  Mandel, Mandic, Mangano, Mansell, Manske, Mantovani, Marchesoni, Marion,
  M\'arka, M\'arka, Markosyan, Maros, Martelli, Martellini, Martin, Martin,
  Martynov, Marx, Mason, Masserot, Massinger, Masso-Reid, Matichard, Matone,
  Mavalvala, Mazumder, Mazzolo, McCarthy, McClelland, McCormick, McGuire,
  McIntyre, McIver, McManus, McWilliams, Meacher, Meadors, Meidam, Melatos,
  Mendell, Mendoza-Gandara, Mercer, Merilh, Merzougui, Meshkov, Messenger,
  Messick, Meyers, Mezzani, Miao, Michel, Middleton, Mikhailov, Milano, Miller,
  Millhouse, Minenkov, Ming, Mirshekari, Mishra, Mitra, Mitrofanov,
  Mitselmakher, Mittleman, Moggi, Mohan, Mohapatra, Montani, Moore, Moore,
  Moraru, Moreno, Morriss, Mossavi, Mours, Mow-Lowry, Mueller, Mueller, Muir,
  Mukherjee, Mukherjee, Mukherjee, Mukund, Mullavey, Munch, Murphy, Murray,
  Mytidis, Nardecchia, Naticchioni, Nayak, Necula, Nedkova, Nelemans, Neri,
  Neunzert, Newton, Nguyen, Nielsen, Nissanke, Nitz, Nocera, Nolting,
  Normandin, Nuttall, Oberling, Ochsner, O'Dell, Oelker, Ogin, Oh, Oh, Ohme,
  Oliver, Oppermann, Oram, O'Reilly, O'Shaughnessy, Ott, Ottaway, Ottens,
  Overmier, Owen, Pai, Pai, Palamos, Palashov, Palomba, Pal-Singh, Pan, Pan,
  Pankow, Pannarale, Pant, Paoletti, Paoli, Papa, Paris, Parker, Pascucci,
  Pasqualetti, Passaquieti, Passuello, Patricelli, Patrick, Pearlstone,
  Pedraza, Pedurand, Pekowsky, Pele, Penn, Perreca, Pfeiffer, Phelps, Piccinni,
  Pichot, Pickenpack, Piergiovanni, Pierro, Pillant, Pinard, Pinto, Pitkin,
  Poeld, Poggiani, Popolizio, Post, Powell, Prasad, Predoi, Premachandra,
  Prestegard, Price, Prijatelj, Principe, Privitera, Prix, Prodi, Prokhorov,
  Puncken, Punturo, Puppo, P\"urrer, Qi, Qin, Quetschke, Quintero,
  Quitzow-James, Raab, Rabeling, Radkins, Raffai, Raja, Rakhmanov, Ramet,
  Rapagnani, Raymond, Razzano, Re, Read, Reed, Regimbau, Rei, Reid, Reitze,
  Rew, Reyes, Ricci, Riles, Robertson, Robie, Robinet, Rocchi, Rolland,
  Rollins, Roma, Romano, Romano, Romanov, Romie, Rosi\ifmmode~\acute{n}\else
  \'{n}\fi{}ska, Rowan, R\"udiger, Ruggi, Ryan, Sachdev, Sadecki, Sadeghian,
  Salconi, Saleem, Salemi, Samajdar, Sammut, Sampson, Sanchez, Sandberg,
  Sandeen, Sanders, Sanders, Sassolas, Sathyaprakash, Saulson, Sauter, Savage,
  Sawadsky, Schale, Schilling, Schmidt, Schmidt, Schnabel, Schofield,
  Sch\"onbeck, Schreiber, Schuette, Schutz, Scott, Scott, Sellers, Sengupta,
  Sentenac, Sequino, Sergeev, Serna, Setyawati, Sevigny, Shaddock, Shaffer,
  Shah, Shahriar, Shaltev, Shao, Shapiro, Shawhan, Sheperd, Shoemaker,
  Shoemaker, Siellez, Siemens, Sigg, Silva, Simakov, Singer, Singer, Singh,
  Singh, Singhal, Sintes, Slagmolen, Smith, Smith, Smith, Smith, Son, Sorazu,
  Sorrentino, Souradeep, Srivastava, Staley, Steinke, Steinlechner,
  Steinlechner, Steinmeyer, Stephens, Stevenson, Stone, Strain, Straniero,
  Stratta, Strauss, Strigin, Sturani, Stuver, Summerscales, Sun, Sutton,
  Swinkels, Szczepa\ifmmode~\acute{n}\else \'{n}\fi{}czyk, Tacca, Talukder,
  Tanner, T\'apai, Tarabrin, Taracchini, Taylor, Theeg, Thirugnanasambandam,
  Thomas, Thomas, Thomas, Thorne, Thorne, Thrane, Tiwari, Tiwari, Tokmakov,
  Tomlinson, Tonelli, Torres, Torrie, T\"oyr\"a, Travasso, Traylor, Trifir\`o,
  Tringali, Trozzo, Tse, Turconi, Tuyenbayev, Ugolini, Unnikrishnan, Urban,
  Usman, Vahlbruch, Vajente, Valdes, Vallisneri, van Bakel, van Beuzekom,
  van~den Brand, Van Den~Broeck, Vander-Hyde, van~der Schaaf, van Heijningen,
  van Veggel, Vardaro, Vass, Vas\'uth, Vaulin, Vecchio, Vedovato, Veitch,
  Veitch, Venkateswara, Verkindt, Vetrano, Vicer\'e, Vinciguerra, Vine, Vinet,
  Vitale, Vo, Vocca, Vorvick, Voss, Vousden, Vyatchanin, Wade, Wade, Wade,
  Waldman, Walker, Wallace, Walsh, Wang, Wang, Wang, Wang, Wang, Ward, Ward,
  Warner, Was, Weaver, Wei, Weinert, Weinstein, Weiss, Welborn, Wen,
  We\ss{}els, Westphal, Wette, Whelan, Whitcomb, White, Whiting, Wiesner,
  Wilkinson, Willems, Williams, Williams, Williamson, Willis, Willke, Wimmer,
  Winkelmann, Winkler, Wipf, Wiseman, Wittel, Woan, Worden, Wright, Wu, Yablon,
  Yakushin, Yam, Yamamoto, Yancey, Yap, Yu, Yvert, Zadro\ifmmode~\dot{z}\else
  \.{z}\fi{}ny, Zangrando, Zanolin, Zendri, Zevin, Zhang, Zhang, Zhang, Zhang,
  Zhao, Zhou, Zhou, Zhu, Zucker, Zuraw, \& Zweizig}]{PhysRevLett.116.061102}
Abbott, B.~P., Abbott, R., Abbott, T.~D., {et~al.} 2016, Phys. Rev. Lett., 116,
  061102

\bibitem[{Abbott {et~al.}(2019)Abbott, Abbott, Abbott, Abraham, Acernese,
  Ackley, Adams, Adhikari, Adya, Affeldt, Agathos, Agatsuma, Aggarwal, Aguiar,
  Aiello, Ain, Ajith, Allen, Allocca, Aloy, Altin, Amato, Ananyeva, Anderson,
  Anderson, Angelova, Antier, Appert, Arai, Araya, Areeda, Ar\`ene, Arnaud,
  Arun, Ascenzi, Ashton, Aston, Astone, Aubin, Aufmuth, AultONeal, Austin,
  Avendano, Avila-Alvarez, Babak, Bacon, Badaracco, Bader, Bae, Baker,
  Baldaccini, Ballardin, Ballmer, Banagiri, Barayoga, Barclay, Barish, Barker,
  Barkett, Barnum, Barone, Barr, Barsotti, Barsuglia, Barta, Bartlett, Bartos,
  Bassiri, Basti, Bawaj, Bayley, Bazzan, B\'ecsy, Bejger, Belahcene, Bell,
  Beniwal, Berger, Bergmann, Bernuzzi, Bero, Berry, Bersanetti, Bertolini,
  Betzwieser, Bhandare, Bidler, Bilenko, Bilgili, Billingsley, Birch, Birney,
  Birnholtz, Biscans, Biscoveanu, Bisht, Bitossi, Bizouard, Blackburn,
  Blackman, Blair, Blair, Blair, Bloemen, Bode, Boer, Boetzel, Bogaert, Bondu,
  Bonilla, Bonnand, Booker, Boom, Booth, Bork, Boschi, Bose, Bossie, Bossilkov,
  Bosveld, Bouffanais, Bozzi, Bradaschia, Brady, Bramley, Branchesi, Brau,
  Briant, Briggs, Brighenti, Brillet, Brinkmann, Brisson, Brockill, Brooks,
  Brown, Brunett, Buikema, Bulik, Bulten, Buonanno, Buskulic,
  Bustamante~Rosell, Buy, Byer, Cabero, Cadonati, Cagnoli, Cahillane,
  Calder\'on~Bustillo, Callister, Calloni, Camp, Campbell, Canepa, Cannon, Cao,
  Cao, Capocasa, Carbognani, Caride, Carney, Carullo, Casanueva~Diaz,
  Casentini, Caudill, Cavagli\`a, Cavalier, Cavalieri, Cella, Cerd\'a-Dur\'an,
  Cerretani, Cesarini, Chaibi, Chakravarti, Chamberlin, Chan, Chao, Charlton,
  Chase, Chassande-Mottin, Chatterjee, Chaturvedi, Chatziioannou, Cheeseboro,
  Chen, Chen, Chen, Cheng, Cheong, Chia, Chincarini, Chiummo, Cho, Cho, Cho,
  Christensen, Chu, Chua, Chung, Chung, Ciani, Ciobanu, Ciolfi, Cipriano,
  Cirone, Clara, Clark, Clearwater, Cleva, Cocchieri, Coccia, Cohadon, Cohen,
  Colgan, Colleoni, Collette, Collins, Cominsky, Constancio, Conti, Cooper,
  Corban, Corbitt, Cordero-Carri\'on, Corley, Cornish, Corsi, Cortese, Costa,
  Cotesta, Coughlin, Coughlin, Coulon, Countryman, Couvares, Covas, Cowan,
  Coward, Cowart, Coyne, Coyne, Creighton, Creighton, Cripe, Croquette,
  Crowder, Cullen, Cumming, Cunningham, Cuoco, Canton, D\'alya, Danilishin,
  D'Antonio, Danzmann, Dasgupta, Da~Silva~Costa, Datrier, Dattilo, Dave,
  Davier, Davis, Daw, DeBra, Deenadayalan, Degallaix, De~Laurentis,
  Del\'eglise, Del~Pozzo, DeMarchi, Demos, Dent, De~Pietri, Derby, De~Rosa,
  De~Rossi, DeSalvo, de~Varona, Dhurandhar, D\'{\i}az, Dietrich, Di~Fiore,
  Di~Giovanni, Di~Girolamo, Di~Lieto, Ding, Di~Pace, Di~Palma, Di~Renzo,
  Dmitriev, Doctor, Donovan, Dooley, Doravari, Dorrington, Downes, Drago,
  Driggers, Du, Ducoin, Dupej, Dwyer, Easter, Edo, Edwards, Effler, Ehrens,
  Eichholz, Eikenberry, Eisenmann, Eisenstein, Essick, Estelles, Estevez,
  Etienne, Etzel, Evans, Evans, Fafone, Fair, Fairhurst, Fan, Farinon, Farr,
  Farr, Fauchon-Jones, Favata, Fays, Fazio, Fee, Feicht, Fejer, Feng,
  Fernandez-Galiana, Ferrante, Ferreira, Ferreira, Ferrini, Fidecaro, Fiori,
  Fiorucci, Fishbach, Fisher, Fishner, Fitz-Axen, Flaminio, Fletcher, Flynn,
  Fong, Font, Forsyth, Fournier, Frasca, Frasconi, Frei, Freise, Frey, Frey,
  Fritschel, Frolov, Fulda, Fyffe, Gabbard, Gadre, Gaebel, Gair, Gammaitoni,
  Ganija, Gaonkar, Garcia, Garc\'{\i}a-Quir\'os, Garufi, Gateley, Gaudio, Gaur,
  Gayathri, Gemme, Genin, Gennai, George, George, Gergely, Germain, Ghonge,
  Ghosh, Ghosh, Ghosh, Giacomazzo, Giaime, Giardina, Giazotto, Gill, Giordano,
  Glover, Godwin, Goetz, Goetz, Goncharov, Gonz\'alez, Gonzalez~Castro,
  Gopakumar, Gorodetsky, Gossan, Gosselin, Gouaty, Grado, Graef, Granata,
  Grant, Gras, Grassia, Gray, Gray, Greco, Green, Green, Gretarsson, Groot,
  Grote, Grunewald, Gruning, Guidi, Gulati, Guo, Gupta, Gupta, Gustafson,
  Gustafson, Haegel, Halim, Hall, Hall, Hamilton, Hammond, Haney, Hanke, Hanks,
  Hanna, Hannam, Hannuksela, Hanson, Hardwick, Haris, Harms, Harry, Harry,
  Haster, Haughian, Hayes, Healy, Heidmann, Heintze, Heitmann, Hello, Hemming,
  Hendry, Heng, Hennig, Heptonstall, Hernandez~Vivanco, Heurs, Hild, Hinderer,
  Hoak, Hochheim, Hofman, Holgado, Holland, Holt, Holz, Hopkins, Horst, Hough,
  Howell, Hoy, Hreibi, Huang, Huerta, Huet, Hughey, Hulko, Husa, Huttner,
  Huynh-Dinh, Idzkowski, Iess, Ingram, Inta, Intini, Irwin, Isa, Isac, Isi,
  Iyer, Izumi, Jacqmin, Jadhav, Jani, Janthalur, Jaranowski, Jenkins, Jiang,
  Johnson, Johnson-McDaniel, Jones, Jones, Jones, Jonker, Ju, Junker,
  Kalaghatgi, Kalogera, Kamai, Kandhasamy, Kang, Kanner, Kapadia, Karki,
  Karvinen, Kashyap, Kasprzack, Katsanevas, Katsavounidis, Katzman, Kaufer,
  Kawabe, Keerthana, K\'ef\'elian, Keitel, Kennedy, Key, Khalili, Khan, Khan,
  Khan, Khan, Khazanov, Khursheed, Kijbunchoo, Kim, Kim, Kim, Kim, Kim, Kim,
  Kimball, King, King, Kinley-Hanlon, Kirchhoff, Kissel, Kleybolte, Klika,
  Klimenko, Knowles, Koch, Koehlenbeck, Koekoek, Koley, Kondrashov, Kontos,
  Koper, Korobko, Korth, Kowalska, Kozak, Kringel, Krishnendu, Kr\'olak, Kuehn,
  Kumar, Kumar, Kumar, Kumar, Kuo, Kutynia, Kwang, Lackey, Lai, Lam, Landry,
  Lane, Lang, Lange, Lantz, Lanza, Lartaux-Vollard, Lasky, Laxen, Lazzarini,
  Lazzaro, Leaci, Leavey, Lecoeuche, Lee, Lee, Lee, Lee, Lee, Lee, Lehmann,
  Lenon, Leroy, Letendre, Levin, Li, Li, Li, Li, Lin, Linde, Linker,
  Littenberg, Liu, Liu, Lo, Lockerbie, London, Longo, Lorenzini, Loriette,
  Lormand, Losurdo, Lough, Lousto, Lovelace, Lower, L\"uck, Lumaca, Lundgren,
  Lynch, Ma, Macas, Macfoy, MacInnis, Macleod, Macquet, Maga\~na Sandoval,
  Maga\~na Zertuche, Magee, Majorana, Maksimovic, Malik, Man, Mandic, Mangano,
  Mansell, Manske, Mantovani, Marchesoni, Marion, M\'arka, M\'arka, Markakis,
  Markosyan, Markowitz, Maros, Marquina, Marsat, Martelli, Martin, Martin,
  Martynov, Mason, Massera, Masserot, Massinger, Masso-Reid, Mastrogiovanni,
  Matas, Matichard, Matone, Mavalvala, Mazumder, McCann, McCarthy, McClelland,
  McCormick, McCuller, McGuire, McIver, McManus, McRae, McWilliams, Meacher,
  Meadors, Mehmet, Mehta, Meidam, Melatos, Mendell, Mercer, Mereni, Merilh,
  Merzougui, Meshkov, Messenger, Messick, Metzdorff, Meyers, Miao, Michel,
  Middleton, Mikhailov, Milano, Miller, Miller, Millhouse, Mills,
  Milovich-Goff, Minazzoli, Minenkov, Mishkin, Mishra, Mistry, Mitra,
  Mitrofanov, Mitselmakher, Mittleman, Mo, Moffa, Mogushi, Mohapatra, Montani,
  Moore, Moraru, Moreno, Morisaki, Mours, Mow-Lowry, Mukherjee, Mukherjee,
  Mukherjee, Mukund, Mullavey, Munch, Mu\~niz, Muratore, Murray, Nagar,
  Nardecchia, Naticchioni, Nayak, Neilson, Nelemans, Nelson, Nery, Neunzert,
  Ng, Ng, Nguyen, Nichols, Nielsen, Nissanke, Nitz, Nocera, North, Nuttall,
  Obergaulinger, Oberling, O'Brien, O'Dea, Ogin, Oh, Oh, Ohme, Ohta, Okada,
  Oliver, Oppermann, Oram, O'Reilly, Ormiston, Ortega, O'Shaughnessy, Ossokine,
  Ottaway, Overmier, Owen, Pace, Pagano, Page, Pai, Pai, Palamos, Palashov,
  Palomba, Pal-Singh, Pan, Pang, Pang, Pankow, Pannarale, Pant, Paoletti,
  Paoli, Papa, Parida, Parker, Pascucci, Pasqualetti, Passaquieti, Passuello,
  Patil, Patricelli, Pearlstone, Pedersen, Pedraza, Pedurand, Pele, Penn,
  Perego, Perez, Perreca, Pfeiffer, Phelps, Phukon, Piccinni, Pichot,
  Piergiovanni, Pillant, Pinard, Pirello, Pitkin, Poggiani, Pong, Ponrathnam,
  Popolizio, Porter, Powell, Prajapati, Prasad, Prasai, Prasanna, Pratten,
  Prestegard, Privitera, Prodi, Prokhorov, Puncken, Punturo, Puppo, P\"urrer,
  Qi, Quetschke, Quinonez, Quintero, Quitzow-James, Raab, Radkins, Radulescu,
  Raffai, Raja, Rajan, Rajbhandari, Rakhmanov, Ramirez, Ramos-Buades, Rana,
  Rao, Rapagnani, Raymond, Razzano, Read, Regimbau, Rei, Reid, Reitze, Ren,
  Ricci, Richardson, Richardson, Ricker, Riemenschneider, Riles, Rizzo,
  Robertson, Robie, Robinet, Rocchi, Rolland, Rollins, Roma, Romanelli, Romano,
  Romel, Romie, Rose, Rosi\ifmmode~\acute{n}\else \'{n}\fi{}ska, Rosofsky,
  Ross, Rowan, R\"udiger, Ruggi, Rutins, Ryan, Sachdev, Sadecki, Sakellariadou,
  Salafia, Salconi, Saleem, Salemi, Samajdar, Sammut, Sanchez, Sanchez,
  Sanchis-Gual, Sandberg, Sanders, Santiago, Sarin, Sassolas, Sathyaprakash,
  Saulson, Sauter, Savage, Schale, Scheel, Scheuer, Schmidt, Schnabel,
  Schofield, Sch\"onbeck, Schreiber, Schulte, Schutz, Schwalbe, Scott, Scott,
  Seidel, Sellers, Sengupta, Sennett, Sentenac, Sequino, Sergeev, Setyawati,
  Shaddock, Shaffer, Shahriar, Shaner, Shao, Sharma, Shawhan, Shen, Shink,
  Shoemaker, Shoemaker, ShyamSundar, Siellez, Sieniawska, Sigg, Silva, Singer,
  Singh, Singhal, Sintes, Sitmukhambetov, Skliris, Slagmolen, Slaven-Blair,
  Smith, Smith, Somala, Son, Sorazu, Sorrentino, Souradeep, Sowell, Spencer,
  Srivastava, Srivastava, Staats, Stachie, Standke, Steer, Steinke,
  Steinlechner, Steinlechner, Steinmeyer, Stevenson, Stocks, Stone, Stops,
  Strain, Stratta, Strigin, Strunk, Sturani, Stuver, Sudhir, Summerscales, Sun,
  Sunil, Suresh, Sutton, Swinkels, Szczepa\ifmmode~\acute{n}\else
  \'{n}\fi{}czyk, Tacca, Tait, Talbot, Talukder, Tanner, T\'apai, Taracchini,
  Tasson, Taylor, Thies, Thomas, Thomas, Thondapu, Thorne, Thrane, Tiwari,
  Tiwari, Tiwari, Toland, Tonelli, Tornasi, Torres-Forn\'e, Torrie, T\"oyr\"a,
  Travasso, Traylor, Tringali, Trovato, Trozzo, Trudeau, Tsang, Tse, Tso,
  Tsukada, Tsuna, Tuyenbayev, Ueno, Ugolini, Unnikrishnan, Urban, Usman,
  Vahlbruch, Vajente, Valdes, van Bakel, van Beuzekom, van~den Brand, Van
  Den~Broeck, Vander-Hyde, van Heijningen, van~der Schaaf, van Veggel, Vardaro,
  Varma, Vass, Vas\'uth, Vecchio, Vedovato, Veitch, Veitch, Venkateswara,
  Venugopalan, Verkindt, Vetrano, Vicer\'e, Viets, Vine, Vinet, Vitale, Vo,
  Vocca, Vorvick, Vyatchanin, Wade, Wade, Wade, Walet, Walker, Wallace, Walsh,
  Wang, Wang, Wang, Wang, Wang, Ward, Warden, Warner, Was, Watchi, Weaver, Wei,
  Weinert, Weinstein, Weiss, Wellmann, Wen, Wessel, We\ss{}els, Westhouse,
  Wette, Whelan, White, Whiting, Whittle, Wilken, Williams, Williamson, Willis,
  Willke, Wimmer, Winkler, Wipf, Wittel, Woan, Woehler, Wofford, Worden,
  Wright, Wu, Wysocki, Xiao, Yamamoto, Yancey, Yang, Yap, Yazback, Yeeles, Yu,
  Yu, Yuen, Yvert, Zadro\ifmmode~\dot{z}\else \.{z}\fi{}ny, Zanolin, Zappa,
  Zelenova, Zendri, Zevin, Zhang, Zhang, Zhang, Zhao, Zhou, Zhou, Zhu,
  Zimmerman, Zlochower, Zucker, \& Zweizig}]{PhysRevX.9.031040}
Abbott, B.~P., Abbott, R., Abbott, T.~D., {et~al.} 2019, Phys. Rev. X, 9,
  031040

\bibitem[{{Abbott} {et~al.}(2019{\natexlab{a}}){Abbott}, {Abbott}, {Abbott},
  {Abraham}, {Acernese}, {Ackley}, {Adams}, {Adhikari}, {Adya}, {Affeldt},
  {Agathos}, {Agatsuma}, {Aggarwal}, {Aguiar}, {Aiello}, {Ain}, {Ajith},
  {Allen}, {Allocca}, {Aloy}, {Altin}, {Amato}, {Ananyeva}, {Anderson},
  {Anderson}, {Angelova}, {Antier}, {Appert}, {Arai}, {Araya}, {Areeda},
  {Ar{\`e}ne}, {Arnaud}, {Arun}, {Ascenzi}, {Ashton}, {Aston}, {Astone},
  {Aubin}, {Aufmuth}, {AultONeal}, {Austin}, {Avendano}, {Avila-Alvarez},
  {Babak}, {Bacon}, {Badaracco}, {Bader}, {Bae}, {Baker}, {Baldaccini},
  {Ballardin}, {Ballmer}, {Banagiri}, {Barayoga}, \&
  {Barclay}}]{2019PhRvD.100b4004A}
{Abbott}, B.~P., {Abbott}, R., {Abbott}, T.~D., {et~al.} 2019{\natexlab{a}},
  \prd, 100, 024004

\bibitem[{{Abbott} {et~al.}(2019{\natexlab{b}}){Abbott}, {Abbott}, {Abbott},
  {Abraham}, {Acernese}, {Ackley}, {Adams}, {Adhikari}, {Adya}, {Affeldt},
  {Agathos}, {Agatsuma}, {Aggarwal}, {Aguiar}, {Aiello}, {Ain}, {Ajith},
  {Allen}, {Allocca}, {Aloy}, {Altin}, {Amato}, {Ananyeva}, {Anderson},
  {Anderson}, {Angelova}, {et~al.}}]{2019PhRvD..99l2002A}
{Abbott}, B.~P., {Abbott}, R., {Abbott}, T.~D., {et~al.} 2019{\natexlab{b}},
  \prd, 99, 122002

\bibitem[{Abbott {et~al.}(2017)Abbott, Abbott, Abbott, Acernese, Ackley, Adams,
  Adams, Addesso, Adhikari, Adya, Affeldt, Afrough, Agarwal, Agathos, Agatsuma,
  Aggarwal, Aguiar, Aiello, Ain, Ajith, Allen, Allen, Allocca, Altin, Amato,
  Ananyeva, Anderson, Anderson, Angelova, Antier, Zucker, \&
  Zweizig}]{PhysRevLett.119.161101}
Abbott, B.~P., Abbott, R., Abbott, T.~D., {et~al.} 2017, Phys. Rev. Lett., 119,
  161101

\bibitem[{{Abbott} {et~al.}(2018{\natexlab{c}}){Abbott}, {Abbott}, {Abbott},
  {Acernese}, {Ackley}, {Adams}, {Adams}, {Addesso}, {Adhikari}, {Adya},
  {Affeldt}, {Afrough}, {Agarwal}, {Agathos}, {Agatsuma}, {Aggarwal}, {Aguiar},
  {Aiello}, {Ain}, {Allen}, {Allen}, {Allocca}, {Altin}, {Amato}, {Ananyeva},
  {Anderson}, {Anderson}, {Angelova}, {Antier}, {Appert}, {Arai}, {Araya},
  {Areeda}, {Arnaud}, {Ascenzi}, {Ashton}, {Ast},
  {et~al.}}]{2018PhRvD..97j2003A}
{Abbott}, B.~P., {Abbott}, R., {Abbott}, T.~D., {et~al.} 2018{\natexlab{c}},
  \prd, 97, 102003

\bibitem[{{Abbott} {et~al.}(2017){Abbott}, {Abbott}, {Abbott}, {Acernese},
  {Ackley}, {Adams}, {Adams}, {Addesso}, {Adhikari}, {Adya}, {Affeldt},
  {Afrough}, {Agarwal}, {Agatsuma}, {Aggarwal}, {Aguiar}, {Aiello}, {Ain},
  {Ajith}, {Allen}, {Allen}, {Allocca}, {Altin}, {Amato}, {Ananyeva},
  {Anderson}, {et~al.}}]{2017PhRvD..96f2002A}
{Abbott}, B.~P., {Abbott}, R., {Abbott}, T.~D., {et~al.} 2017, \prd, 96, 062002

\bibitem[{Abbott {et~al.}(2017)Abbott, Abbott, Abbott, Acernese, Ackley, Adams,
  Adams, Addesso, Adhikari, Adya, Affeldt, Afrough, Agarwal, Agatsuma,
  Aggarwal, Aguiar, Aiello, Ain, Allen, Allen, Allocca, Altin, Amato, Ananyeva,
  Anderson, Anderson, Antier, Appert, Arai, Araya, Areeda, Arnaud, Ascenzi,
  Ashton, Ast, Aston, Astone, Aufmuth, Aulbert, AultONeal, Avila-Alvarez,
  Babak, Bacon, Bader, Bae, Baker, Baldaccini, Ballardin, Ballmer, Banagiri,
  Barayoga, Barclay, Barish, Barker, Barone, Barr, Barsotti, Barsuglia, Barta,
  Bartlett, Bartos, Bassiri, Basti, Batch, Baune, Bawaj, Bazzan, B\'ecsy, Beer,
  Bejger, Belahcene, Bell, Berger, Bergmann, Berry, Bersanetti, Bertolini,
  Betzwieser, Bhagwat, Bhandare, Bilenko, Billingsley, Billman, Birch, Birney,
  Birnholtz, Biscans, Bisht, Bitossi, Biwer, Bizouard, Blackburn, Blackman,
  Blair, Blair, Blair, Bloemen, Bock, Bode, Boer, Bogaert, Bohe, Bondu,
  Bonnand, Boom, Bork, Boschi, Bose, Bouffanais, Bozzi, Bradaschia, Brady,
  Braginsky, Branchesi, Brau, Briant, Brillet, Brinkmann, Brisson, Brockill,
  Broida, Brooks, Brown, Brown, Brown, Brunett, Buchanan, Buikema, Bulik,
  Bulten, Buonanno, Buskulic, Buy, Byer, Cabero, Cadonati, Cagnoli, Cahillane,
  Calder\'on~Bustillo, Callister, Calloni, Camp, Canizares, Cannon, Cao, Cao,
  Capano, Capocasa, Carbognani, Caride, Carney, Casanueva~Diaz, Casentini,
  Caudill, Cavagli\`a, Cavalier, Cavalieri, Cella, Cepeda, Cerboni~Baiardi,
  Cerretani, Cesarini, Chamberlin, Chan, Chao, Charlton, Chassande-Mottin,
  Chatterjee, Cheeseboro, Chen, Chen, Cheng, Chincarini, Chiummo, Chmiel, Cho,
  Cho, Chow, Christensen, Chu, Chua, Chua, Chung, Chung, Ciani, Ciolfi,
  Cirelli, Cirone, Clara, Clark, Cleva, Cocchieri, Coccia, Cohadon, Colla,
  Collette, Cominsky, Constancio, Conti, Cooper, Corban, Corbitt, Corley,
  Cornish, Corsi, Cortese, Costa, Coughlin, Coughlin, Coulon, Countryman,
  Couvares, Covas, Cowan, Coward, Cowart, Coyne, Coyne, Creighton, Creighton,
  Cripe, Crowder, Cullen, Cumming, Cunningham, Cuoco, Dal~Canton, Danilishin,
  D'Antonio, Danzmann, Dasgupta, Da~Silva~Costa, Dattilo, Dave, Davier, Davis,
  Daw, Day, De, DeBra, Deelman, Degallaix, De~Laurentis, Del\'eglise,
  Del~Pozzo, Denker, Dent, Dergachev, De~Rosa, DeRosa, DeSalvo, Devenson,
  Devine, Dhurandhar, D\'{\i}az, Di~Fiore, Di~Giovanni, Di~Girolamo, Di~Lieto,
  Di~Pace, Di~Palma, Di~Renzo, Doctor, Dolique, Donovan, Dooley, Doravari,
  Dorrington, Douglas, Dovale~\'Alvarez, Downes, Drago, Drever, Driggers, Du,
  Ducrot, Duncan, Dwyer, Edo, Edwards, Effler, Eggenstein, Ehrens, Eichholz,
  Eikenberry, Eisenstein, Essick, Etienne, Etzel, Evans, Evans, Factourovich,
  Fafone, Fair, Fairhurst, Fan, Farinon, Farr, Farr, Fauchon-Jones, Favata,
  Fays, Fehrmann, Feicht, Fejer, Fernandez-Galiana, Ferrante, Ferreira,
  Ferrini, Fidecaro, Fiori, Fiorucci, Fisher, Flaminio, Fletcher, Fong,
  Forsyth, Forsyth, Fournier, Frasca, Frasconi, Frei, Freise, Frey, Frey,
  Fries, Fritschel, Frolov, Fulda, Fyffe, Gabbard, Gabel, Gadre, Gaebel, Gair,
  Gammaitoni, Ganija, Gaonkar, Garufi, Gaudio, Gaur, Gayathri, Gehrels, Gemme,
  Genin, Gennai, George, George, Gergely, Germain, Ghonge, Ghosh, Ghosh, Ghosh,
  Giaime, Giardina, Giazotto, Gill, Glover, Goetz, Goetz, Gomes, Gonz\'alez,
  Gonzalez~Castro, Gopakumar, Gorodetsky, Gossan, Gosselin, Gouaty, Grado,
  Graef, Granata, Grant, Gras, Gray, Greco, Green, Groot, Grote, Grunewald,
  Gruning, Guidi, Guo, Gupta, Gupta, Gushwa, Gustafson, Gustafson, Hall, Hall,
  Hammond, Haney, Hanke, Hanks, Hanna, Hannuksela, Hanson, Hardwick, Harms,
  Harry, Harry, Hart, Haster, Haughian, Healy, Heidmann, Heintze, Heitmann,
  Hello, Hemming, Hendry, Heng, Hennig, Henry, Heptonstall, Heurs, Hild, Hoak,
  Hofman, Holt, Holz, Hopkins, Horst, Hough, Houston, Howell, Hu, Huerta, Huet,
  Hughey, Husa, Huttner, Huynh-Dinh, Indik, Ingram, Inta, Intini, Isa, Isac,
  Isi, Iyer, Izumi, Jacqmin, Jani, Jaranowski, Jawahar, Jim\'enez-Forteza,
  Johnson, Jones, Jones, Jonker, Ju, Junker, Kalaghatgi, Kalogera, Kandhasamy,
  Kang, Kanner, Karki, Karvinen, Kasprzack, Katolik, Katsavounidis, Katzman,
  Kaufer, Kawabe, K\'ef\'elian, Keitel, Kemball, Kennedy, Kent, Key, Khalili,
  Khan, Khan, Khan, Khazanov, Kijbunchoo, Kim, Kim, Kim, Kim, Kim, Kimbrell,
  King, King, Kirchhoff, Kissel, Kleybolte, Klimenko, Koch, Koehlenbeck, Koley,
  Kondrashov, Kontos, Korobko, Korth, Kowalska, Kozak, Kr\"amer, Kringel,
  Krishnan, Kr\'olak, Kuehn, Kumar, Kumar, Kumar, Kuo, Kutynia, Kwang, Lackey,
  Lai, Landry, Lang, Lange, Lantz, Lanza, Lartaux-Vollard, Lasky, Laxen,
  Lazzarini, Lazzaro, Leaci, Leavey, Lee, Lee, Lee, Lee, Lee, Lehmann, Lenon,
  Leonardi, Leroy, Letendre, Levin, Li, Libson, Littenberg, Liu, Lo, Lockerbie,
  London, Lord, Lorenzini, Loriette, Lormand, Losurdo, Lough, Lovelace, L\"uck,
  Lumaca, Lundgren, Lynch, Ma, Macfoy, Machenschalk, MacInnis, Macleod,
  Maga\~na Hernandez, Maga\~na Sandoval, Maga\~na Zertuche, Magee, Majorana,
  Maksimovic, Man, Mandic, Mangano, Mansell, Manske, Mantovani, Marchesoni,
  Marion, M\'arka, M\'arka, Markakis, Markosyan, Maros, Martelli, Martellini,
  Martin, Martynov, Mason, Masserot, Massinger, Masso-Reid, Mastrogiovanni,
  Matas, Matichard, Matone, Mavalvala, Mayani, Mazumder, McCarthy, McClelland,
  McCormick, McCuller, McGuire, McIntyre, McIver, McManus, McRae, McWilliams,
  Meacher, Meadors, Meidam, Mejuto-Villa, Melatos, Mendell, Mercer, Merilh,
  Merzougui, Meshkov, Messenger, Messick, Metzdorff, Meyers, Mezzani, Miao,
  Michel, Middleton, Mikhailov, Milano, Miller, Miller, Miller, Miller,
  Millhouse, Minazzoli, Minenkov, Ming, Mishra, Mitra, Mitrofanov,
  Mitselmakher, Mittleman, Moggi, Mohan, Mohapatra, Montani, Moore, Moore,
  Moraru, Moreno, Morriss, Mours, Mow-Lowry, Mueller, Muir, Mukherjee,
  Mukherjee, Mukherjee, Mukund, Mullavey, Munch, Muniz, Murray, Napier,
  Nardecchia, Naticchioni, Nayak, Nelemans, Nelson, Neri, Nery, Neunzert,
  Newport, Newton, Ng, Nguyen, Nichols, Nielsen, Nissanke, Nitz, Noack, Nocera,
  Nolting, Normandin, Nuttall, Oberling, Ochsner, Oelker, Ogin, Oh, Oh, Ohme,
  Oliver, Oppermann, Oram, O'Reilly, Ormiston, Ortega, O'Shaughnessy, Ottaway,
  Overmier, Owen, Pace, Page, Page, Pai, Pai, Palamos, Palashov, Palomba,
  Pal-Singh, Pan, Pang, Pang, Pankow, Pannarale, Pant, Paoletti, Paoli, Papa,
  Paris, Parker, Pascucci, Pasqualetti, Passaquieti, Passuello, Patricelli,
  Pearlstone, Pedraza, Pedurand, Pekowsky, Pele, Penn, Perez, Perreca, Perri,
  Pfeiffer, Phelps, Piccinni, Pichot, Piergiovanni, Pierro, Pillant, Pinard,
  Pinto, Pitkin, Poggiani, Popolizio, Porter, Post, Powell, Prasad, Pratt,
  Predoi, Prestegard, Prijatelj, Principe, Privitera, Prix, Prodi, Prokhorov,
  Puncken, Punturo, Puppo, P\"urrer, Qi, Qin, Qiu, Quetschke, Quintero,
  Quitzow-James, Raab, Rabeling, Radkins, Raffai, Raja, Rajan, Rakhmanov,
  Ramirez, Rapagnani, Raymond, Razzano, Read, Regimbau, Rei, Reid, Reitze, Rew,
  Reyes, Ricci, Ricker, Rieger, Riles, Rizzo, Robertson, Robie, Robinet,
  Rocchi, Rolland, Rollins, Roma, Romano, Romel, Romie,
  Rosi\ifmmode~\acute{n}\else \'{n}\fi{}ska, Ross, Rowan, R\"udiger, Ruggi,
  Ryan, Rynge, Sachdev, Sadecki, Sadeghian, Sakellariadou, Salconi, Saleem,
  Salemi, Samajdar, Sammut, Sampson, Sanchez, Sandberg, Sandeen, Sanders,
  Sassolas, Sathyaprakash, Saulson, Sauter, Savage, Sawadsky, Schale, Scheuer,
  Schmidt, Schmidt, Schmidt, Schnabel, Schofield, Sch\"onbeck, Schreiber,
  Schuette, Schulte, Schutz, Schwalbe, Scott, Scott, Seidel, Sellers, Sengupta,
  Sentenac, Sequino, Sergeev, Shaddock, Shaffer, Shah, Shahriar, Shao, Shapiro,
  Shawhan, Sheperd, Shoemaker, Shoemaker, Siellez, Siemens, Sieniawska, Sigg,
  Silva, Singer, Singer, Singh, Singh, Singhal, Sintes, Slagmolen, Smith,
  Smith, Smith, Son, Sonnenberg, Sorazu, Sorrentino, Souradeep, Spencer,
  Srivastava, Staley, Steinke, Steinlechner, Steinlechner, Steinmeyer,
  Stephens, Stone, Strain, Stratta, Strigin, Sturani, Stuver, Summerscales,
  Sun, Sunil, Sutton, Swinkels, Szczepa\ifmmode~\acute{n}\else \'{n}\fi{}czyk,
  Tacca, Talukder, Tanner, T\'apai, Taracchini, Taylor, Taylor, Theeg, Thomas,
  Thomas, Thomas, Thorne, Thorne, Thrane, Tiwari, Tiwari, Tokmakov, Toland,
  Tonelli, Tornasi, Torrie, T\"oyr\"a, Travasso, Traylor, Trifir\`o, Trinastic,
  Tringali, Trozzo, Tsang, Tse, Tso, Tuyenbayev, Ueno, Ugolini, Unnikrishnan,
  Urban, Usman, Vahi, Vahlbruch, Vajente, Valdes, Vallisneri, van Bakel, van
  Beuzekom, van~den Brand, Van Den~Broeck, Vander-Hyde, van~der Schaaf, van
  Heijningen, van Veggel, Vardaro, Varma, Vass, Vas\'uth, Vecchio, Vedovato,
  Veitch, Veitch, Venkateswara, Venugopalan, Verkindt, Vetrano, Vicer\'e,
  Viets, Vinciguerra, Vine, Vinet, Vitale, Vo, Vocca, Vorvick, Voss, Vousden,
  Vyatchanin, Wade, Wade, Wade, Walet, Walker, Wallace, Walsh, Wang, Wang,
  Wang, Wang, Wang, Wang, Ward, Warner, Was, Watchi, Weaver, Wei, Weinert,
  Weinstein, Weiss, Wen, Wessel, We\ss{}els, Westphal, Wette, Whelan, Whiting,
  Whittle, Williams, Williams, Williamson, Willis, Willke, Wimmer, Winkler,
  Wipf, Wittel, Woan, Woehler, Wofford, Wong, Worden, Wright, Wu, Wu, Yam,
  Yamamoto, Yancey, Yap, Yu, Yu, Yvert, Zadro\ifmmode~\dot{z}\else
  \.{z}\fi{}ny, Zanolin, Zelenova, Zendri, Zevin, Zhang, Zhang, Zhang, Zhang,
  Zhao, Zhou, Zhou, Zhu, Zhu, Zucker, Zweizig, \&
  Anderson}]{PhysRevD.96.122004}
Abbott, B.~P., Abbott, R., Abbott, T.~D., {et~al.} 2017, Phys. Rev. D, 96,
  122004

\bibitem[{{Abbott} {et~al.}(2017){Abbott}, {Abbott}, {Abbott}, {Acernese},
  {Ackley}, {Adams}, {Adams}, {Addesso}, {Adhikari}, {Adya}, \&
  et~al.}]{2017ApJ...848L..12A}
{Abbott}, B.~P., {Abbott}, R., {Abbott}, T.~D., {et~al.} 2017, \apjl, 848, L12

\bibitem[{{Abbott} {et~al.}(2020){Abbott}, {Abbott}, {Abraham}, {Acernese},
  {Ackley}, {Adams}, {Adams}, {Adhikari}, {Adya}, {Affeldt}, {Agathos},
  {Agatsuma}, {Aggarwal}, {Aguiar}, {Aiello}, {Ain}, {Ajith}, {Akcay}, {Allen},
  {Allocca}, {Altin}, {Amato}, {Anand}, {Ananyeva}, {Anderson}, {Anderson},
  {Angelova}, {Ansoldi}, {Antelis}, {Antier}, {Appert}, {Arai}, {Araya},
  {Areeda}, {Ar{\`e}ne}, {Arnaud}, {Aronson}, {Arun}, {Asali},
  {et~al.}}]{2020arXiv201014527A}
{Abbott}, R., {Abbott}, T.~D., {Abraham}, S., {et~al.} 2020, arXiv e-prints,
  arXiv:2010.14527

\bibitem[{{Acernese} {et~al.}(2015){Acernese}, {Agathos}, {Agatsuma}, {Aisa},
  {Allemandou}, {Allocca}, {Amarni}, {Astone}, {Balestri}, {Ballardin},
  {Barone}, {Baronick}, {Barsuglia}, {Basti}, {Basti}, {Bauer}, {Bavigadda},
  {Bejger}, {Beker}, {Belczynski}, {Bersanetti}, {Bertolini}, {Bitossi},
  {Bizouard}, {Bloemen}, {Blom}, {Boer}, {Bogaert}, {Bondi}, {Bondu},
  {Bonelli}, {Bonnand}, {Boschi}, {Bosi}, {Bouedo}, {Bradaschia}, {Branchesi},
  {Briant}, {Brillet}, {Brisson}, {Bulik}, {Bulten}, {Buskulic}, {Buy},
  {Cagnoli}, {Calloni}, {Campeggi}, {Canuel}, {Carbognani}, {Cavalier},
  {Cavalieri}, {Cella}, {Cesarini}, {Chassande-Mottin}, {Chincarini},
  {Chiummo}, {Chua}, {Cleva}, {Coccia}, {Cohadon}, {Colla}, {Colombini},
  {Conte}, {Coulon}, {Cuoco}, {Dalmaz}, {D'Antonio}, {Dattilo}, {Davier},
  {Day}, {Debreczeni}, {Degallaix}, {Del{\'e}glise}, {Del Pozzo}, {Dereli}, {De
  Rosa}, {Di Fiore}, {Di Lieto}, {Di Virgilio}, {Doets}, {Dolique}, {Drago},
  {Ducrot}, {Endr{\H{o}}czi}, {Fafone}, {Farinon}, {Ferrante}, {Ferrini},
  {Fidecaro}, {Fiori}, {Flaminio}, {Fournier}, {Franco}, {Frasca}, {Frasconi},
  {Gammaitoni}, {Garufi}, {Gaspard}, {Gatto}, {Gemme}, {Gendre}, {Genin},
  {Gennai}, {Ghosh}, {Giacobone}, {Giazotto}, {Gouaty}, {Granata}, {Greco},
  {Groot}, {Guidi}, {Harms}, {Heidmann}, {Heitmann}, {Hello}, {Hemming},
  {Hennes}, {Hofman}, {Jaranowski}, {Jonker}, {Kasprzack}, {K{\'e}f{\'e}lian},
  {Kowalska}, {Kraan}, {Kr{\'o}lak}, {Kutynia}, {Lazzaro}, {Leonardi}, {Leroy},
  {Letendre}, {Li}, {Lieunard}, {Lorenzini}, {Loriette}, {Losurdo},
  {Magazz{\`u}}, {Majorana}, {Maksimovic}, {Malvezzi}, {Man}, {Mangano},
  {Mantovani}, {Marchesoni}, {Marion}, {Marque}, {Martelli}, {Martellini},
  {Masserot}, {Meacher}, {Meidam}, {Mezzani}, {Michel}, {Milano}, {Minenkov},
  {Moggi}, {Mohan}, {Montani}, {Morgado}, {Mours}, {Mul}, {Nagy}, {Nardecchia},
  {Naticchioni}, {Nelemans}, {Neri}, {Neri}, {Nocera}, {Pacaud}, {Palomba},
  {Paoletti}, {Paoli}, {Pasqualetti}, {Passaquieti}, {Passuello}, {Perciballi},
  {Petit}, {Pichot}, {Piergiovanni}, {Pillant}, {Piluso}, {Pinard}, {Poggiani},
  {Prijatelj}, {Prodi}, {Punturo}, {Puppo}, {Rabeling}, {R{\'a}cz},
  {Rapagnani}, {Razzano}, {Re}, {Regimbau}, {Ricci}, {Robinet}, {Rocchi},
  {Rolland}, {Romano}, {Rosi{\'n}ska}, {Ruggi}, {Saracco}, {Sassolas},
  {Schimmel}, {Sentenac}, {Sequino}, {Shah}, {Siellez}, {Straniero},
  {Swinkels}, {Tacca}, {Tonelli}, {Travasso}, {Turconi}, {Vajente}, {van
  Bakel}, {van Beuzekom}, {van den Brand}, {Van Den Broeck}, {van der Sluys},
  {van Heijningen}, {Vas{\'u}th}, {Vedovato}, {Veitch}, {Verkindt}, {Vetrano},
  {Vicer{\'e}}, {Vinet}, {Visser}, {Vocca}, {Ward}, {Was}, {Wei}, {Yvert},
  {Zadro {\.z}ny}, \& {Zendri}}]{2015CQGra..32b4001A}
{Acernese}, F., {Agathos}, M., {Agatsuma}, K., {et~al.} 2015, Classical and
  Quantum Gravity, 32, 024001

\bibitem[{{Andersson} {et~al.}(2011){Andersson}, {Ferrari}, {Jones},
  {Kokkotas}, {Krishnan}, {Read}, {Rezzolla}, \& {Zink}}]{2011GReGr..43..409A}
{Andersson}, N., {Ferrari}, V., {Jones}, D.~I., {et~al.} 2011, General
  Relativity and Gravitation, 43, 409

\bibitem[{Andersson {et~al.}(1999)Andersson, Kokkotas, \&
  Stergioulas}]{Andersson_1999}
Andersson, N., Kokkotas, K.~D., \& Stergioulas, N. 1999, The Astrophysical
  Journal, 516, 307

\bibitem[{Antonucci {et~al.}(2008)Antonucci, Astone, Antonio, Frasca, \&
  Palomba}]{Antonucci_2008}
Antonucci, F., Astone, P., Antonio, S.~D., Frasca, S., \& Palomba, C. 2008,
  Classical and Quantum Gravity, 25, 184015

\bibitem[{Astone {et~al.}(2014)Astone, Colla, D'Antonio, Frasca, \&
  Palomba}]{PhysRevD.90.042002}
Astone, P., Colla, A., D'Antonio, S., Frasca, S., \& Palomba, C. 2014, Phys.
  Rev. D, 90, 042002

\bibitem[{Astone {et~al.}(2010)Astone, D'Antonio, Frasca, \&
  Palomba}]{Astone_2010}
Astone, P., D'Antonio, S., Frasca, S., \& Palomba, C. 2010, Classical and
  Quantum Gravity, 27, 194016

\bibitem[{{Belczynski} {et~al.}(2008){Belczynski}, {Kalogera}, {Rasio}, {Taam},
  {Zezas}, {Bulik}, {Maccarone}, \& {Ivanova}}]{2008ApJS..174..223B}
{Belczynski}, K., {Kalogera}, V., {Rasio}, F.~A., {et~al.} 2008, \apjs, 174,
  223

\bibitem[{{Bhattacharya} \& {van den Heuvel}(1991)}]{1991PhR...203....1B}
{Bhattacharya}, D. \& {van den Heuvel}, E.~P.~J. 1991, \physrep, 203, 1

\bibitem[{Bildsten(1998)}]{Bildsten_1998}
Bildsten, L. 1998, The Astrophysical Journal, 501, L89

\bibitem[{{Bonazzola} \& {Gourgoulhon}(1996)}]{1996A&A...312..675B}
{Bonazzola}, S. \& {Gourgoulhon}, E. 1996, \aap, 312, 675

\bibitem[{{Caride} {et~al.}(2019){Caride}, {Inta}, {Owen}, \& {Rajbhand
  ari}}]{2019PhRvD.100f4013C}
{Caride}, S., {Inta}, R., {Owen}, B.~J., \& {Rajbhand ari}, B. 2019, \prd, 100,
  064013

\bibitem[{{Chau}(1970)}]{1970Natur.228..655C}
{Chau}, W.~Y. 1970, \nat, 228, 655

\bibitem[{{Cie{\'s}lar} {et~al.}(2020){Cie{\'s}lar}, {Bulik}, \&
  {Os{\l}owski}}]{2020MNRAS.492.4043C}
{Cie{\'s}lar}, M., {Bulik}, T., \& {Os{\l}owski}, S. 2020, \mnras, 492, 4043

\bibitem[{{Dergachev} \& {Papa}(2019)}]{2019PhRvL.123j1101D}
{Dergachev}, V. \& {Papa}, M.~A. 2019, \prl, 123, 101101

\bibitem[{{Dergachev} \& {Papa}(2020)}]{2020PhRvL.125q1101D}
{Dergachev}, V. \& {Papa}, M.~A. 2020, \prl, 125, 171101

\bibitem[{{Dreissigacker} \& {Prix}(2020)}]{2020PhRvD.102b2005D}
{Dreissigacker}, C. \& {Prix}, R. 2020, \prd, 102, 022005

\bibitem[{Dreissigacker {et~al.}(2018)Dreissigacker, Prix, \&
  Wette}]{PhysRevD.98.084058}
Dreissigacker, C., Prix, R., \& Wette, K. 2018, Phys. Rev. D, 98, 084058

\bibitem[{Dupuis \& Woan(2005)}]{PhysRevD.72.102002}
Dupuis, R.~J. \& Woan, G. 2005, Phys. Rev. D, 72, 102002

\bibitem[{Hild {et~al.}(2011)Hild, Abernathy, Acernese, Amaro-Seoane,
  Andersson, Arun, Barone, Barr, Barsuglia, Beker, Beveridge, Birindelli, Bose,
  Bosi, Braccini, Bradaschia, Bulik, Calloni, Cella, Mottin, Chelkowski,
  Chincarini, Clark, Coccia, Colacino, Colas, Cumming, Cunningham, Cuoco,
  Danilishin, Danzmann, Salvo, Dent, Rosa, Fiore, Virgilio, Doets, Fafone,
  Falferi, Flaminio, Franc, Frasconi, Freise, Friedrich, Fulda, Gair, Gemme,
  Genin, Gennai, Giazotto, Glampedakis, Gräf, Granata, Grote, Guidi,
  Gurkovsky, Hammond, Hannam, Harms, Heinert, Hendry, Heng, Hennes, Hough,
  Husa, Huttner, Jones, Khalili, Kokeyama, Kokkotas, Krishnan, Li, Lorenzini,
  Lück, Majorana, Mandel, Mandic, Mantovani, Martin, Michel, Minenkov,
  Morgado, Mosca, Mours, Müller{\textendash}Ebhardt, Murray, Nawrodt, Nelson,
  Oshaughnessy, Ott, Palomba, Paoli, Parguez, Pasqualetti, Passaquieti,
  Passuello, Pinard, Plastino, Poggiani, Popolizio, Prato, Punturo, Puppo,
  Rabeling, Rapagnani, Read, Regimbau, Rehbein, Reid, Ricci, Richard, Rocchi,
  Rowan, Rüdiger, Santamar{\'{\i}}a, Sassolas, Sathyaprakash, Schnabel,
  Schwarz, Seidel, Sintes, Somiya, Speirits, Strain, Strigin, Sutton, Tarabrin,
  Thüring, van~den Brand, van Veggel, van~den Broeck, Vecchio, Veitch,
  Vetrano, Vicere, Vyatchanin, Willke, Woan, \& Yamamoto}]{Hild_2011}
Hild, S., Abernathy, M., Acernese, F., {et~al.} 2011, Classical and Quantum
  Gravity, 28, 094013

\bibitem[{{Hobbs} {et~al.}(2005){Hobbs}, {Lorimer}, {Lyne}, \&
  {Kramer}}]{2005MNRAS.360..974H}
{Hobbs}, G., {Lorimer}, D.~R., {Lyne}, A.~G., \& {Kramer}, M. 2005, \mnras,
  360, 974

\bibitem[{Jaranowski \& Krolak(2009)}]{jaranowski_analysis_2009}
Jaranowski, P. \& Krolak, A. 2009, Analysis of {Gravitational}-{Wave} {Data},
  Cambridge {Monographs} on {Particle} {Physics}, {Nuclear} {Physics} and
  {Cosmology} No.~29 (Cambridge University Press)

\bibitem[{{Jaranowski} {et~al.}(1998){Jaranowski}, {Kr{\'o}lak}, \&
  {Schutz}}]{1998PhRvD..58f3001J}
{Jaranowski}, P., {Kr{\'o}lak}, A., \& {Schutz}, B.~F. 1998, \prd, 58, 063001

\bibitem[{{Kiziltan} \& {Thorsett}(2009)}]{2009ApJ...693L.109K}
{Kiziltan}, B. \& {Thorsett}, S.~E. 2009, \apjl, 693, L109

\bibitem[{Knispel \& Allen(2008)}]{PhysRevD.78.044031}
Knispel, B. \& Allen, B. 2008, Phys. Rev. D, 78, 044031

\bibitem[{Krishnan {et~al.}(2004)Krishnan, Sintes, Papa, Schutz, Frasca, \&
  Palomba}]{PhysRevD.70.082001}
Krishnan, B., Sintes, A.~M., Papa, M.~A., {et~al.} 2004, Phys. Rev. D, 70,
  082001

\bibitem[{{Lasky}(2015)}]{2015PASA...32...34L}
{Lasky}, P.~D. 2015, \pasa, 32, e034

\bibitem[{{Lindblom} {et~al.}(1998){Lindblom}, {Owen}, \&
  {Morsink}}]{1998PhRvL..80.4843L}
{Lindblom}, L., {Owen}, B.~J., \& {Morsink}, S.~M. 1998, \prl, 80, 4843

\bibitem[{Maggiore {et~al.}(2020)Maggiore, Broeck, Bartolo, Belgacem, Bertacca,
  Bizouard, Branchesi, Clesse, Foffa, Garc{\'{\i}}a-Bellido, Grimm, Harms,
  Hinderer, Matarrese, Palomba, Peloso, Ricciardone, \&
  Sakellariadou}]{Maggiore_2020}
Maggiore, M., Broeck, C. V.~D., Bartolo, N., {et~al.} 2020, Journal of
  Cosmology and Astroparticle Physics, 2020, 050

\bibitem[{{Manchester} {et~al.}(2005){Manchester}, {Hobbs}, {Teoh}, \&
  {Hobbs}}]{2005AJ....129.1993M}
{Manchester}, R.~N., {Hobbs}, G.~B., {Teoh}, A., \& {Hobbs}, M. 2005, \aj, 129,
  1993

\bibitem[{{Melosh}(1969)}]{1969Natur.224..781M}
{Melosh}, H.~J. 1969, \nat, 224, 781

\bibitem[{Miller {et~al.}(2018)Miller, Astone, D'Antonio, Frasca, Intini,
  La~Rosa, Leaci, Mastrogiovanni, Muciaccia, Palomba, Piccinni, Singhal, \&
  Whiting}]{PhysRevD.98.102004}
Miller, A., Astone, P., D'Antonio, S., {et~al.} 2018, Phys. Rev. D, 98, 102004

\bibitem[{{Ostriker} \& {Gunn}(1969)}]{1969ApJ...157.1395O}
{Ostriker}, J.~P. \& {Gunn}, J.~E. 1969, \apj, 157, 1395

\bibitem[{Owen {et~al.}(1998)Owen, Lindblom, Cutler, Schutz, Vecchio, \&
  Andersson}]{PhysRevD.58.084020}
Owen, B.~J., Lindblom, L., Cutler, C., {et~al.} 1998, Phys. Rev. D, 58, 084020

\bibitem[{{Palomba}(2005)}]{2005MNRAS.359.1150P}
{Palomba}, C. 2005, \mnras, 359, 1150

\bibitem[{Piccinni {et~al.}(2018)Piccinni, Astone, D'Antonio, Frasca, Intini,
  Leaci, Mastrogiovanni, Miller, Palomba, \& Singhal}]{Piccinni_2018}
Piccinni, O.~J., Astone, P., D'Antonio, S., {et~al.} 2018, Classical and
  Quantum Gravity, 36, 015008

\bibitem[{{Press} \& {Thorne}(1972)}]{1972ARA&A..10..335P}
{Press}, W.~H. \& {Thorne}, K.~S. 1972, \araa, 10, 335

\bibitem[{{Punturo} {et~al.}(2010){Punturo}, {Abernathy}, {Acernese}, {Allen},
  {Andersson}, {Arun}, {Barone}, {Barr}, {Barsuglia}, {Beker}, {Beveridge},
  {Birindelli}, {Bose}, {Bosi}, {Braccini}, {Bradaschia}, {Bulik}, {Calloni},
  {Cella}, {Chassande Mottin}, {Chelkowski}, {Chincarini}, {Clark}, {Coccia},
  {Colacino}, {Colas}, {Cumming}, {Cunningham}, {Cuoco}, {Danilishin},
  {Danzmann}, {De Luca}, {De Salvo}, {Dent}, {De Rosa}, {Di Fiore}, {Di
  Virgilio}, {Doets}, {Fafone}, {Falferi}, {Flaminio}, {Franc}, {Frasconi},
  {Freise}, {Fulda}, {Gair}, {Gemme}, {Gennai}, {Giazotto}, {Glampedakis},
  {Granata}, {Grote}, {Guidi}, {Hammond}, {Hannam}, {Harms}, {Heinert},
  {Hendry}, {Heng}, {Hennes}, {Hild}, {Hough}, {Husa}, {Huttner}, {Jones},
  {Khalili}, {Kokeyama}, {Kokkotas}, {Krishnan}, {Lorenzini}, {L{\"u}ck},
  {Majorana}, {Mandel}, {Mandic}, {Martin}, {Michel}, {Minenkov}, {Morgado},
  {Mosca}, {Mours}, {M{\"u}ller─Ebhardt}, {Murray}, {Nawrodt}, {Nelson},
  {Oshaughnessy}, {Ott}, {Palomba}, {Paoli}, {Parguez}, {Pasqualetti},
  {Passaquieti}, {Passuello}, {Pinard}, {Poggiani}, {Popolizio}, {Prato},
  {Puppo}, {Rabeling}, {Rapagnani}, {Read}, {Regimbau}, {Rehbein}, {Reid},
  {Rezzolla}, {Ricci}, {Richard}, {Rocchi}, {Rowan}, {R{\"u}diger}, {Sassolas},
  {Sathyaprakash}, {Schnabel}, {Schwarz}, {Seidel}, {Sintes}, {Somiya},
  {Speirits}, {Strain}, {Strigin}, {Sutton}, {Tarabrin}, {Th{\"u}ring}, {van
  den Brand}, {van Leewen}, {van Veggel}, {van den Broeck}, {Vecchio},
  {Veitch}, {Vetrano}, {Vicere}, {Vyatchanin}, {Willke}, {Woan}, {Wolfango}, \&
  {Yamamoto}}]{2010CQGra..27s4002P}
{Punturo}, M., {Abernathy}, M., {Acernese}, F., {et~al.} 2010, Classical and
  Quantum Gravity, 27, 194002

\bibitem[{{Regimbau} \& {de Freitas Pacheco}(2000)}]{2000A&A...359..242R}
{Regimbau}, T. \& {de Freitas Pacheco}, J.~A. 2000, \aap, 359, 242

\bibitem[{{Riles}(2017)}]{2017MPLA...3230035R}
{Riles}, K. 2017, Modern Physics Letters A, 32, 1730035

\bibitem[{Rudak \& Ritter(1994)}]{10.1093/mnras/267.3.513}
Rudak, B. \& Ritter, H. 1994, Monthly Notices of the Royal Astronomical
  Society, 267, 513

\bibitem[{{Sieniawska} \& {Bejger}(2019)}]{2019Univ....5..217S}
{Sieniawska}, M. \& {Bejger}, M. 2019, Universe, 5, 217

\bibitem[{{Steltner} {et~al.}(2020){Steltner}, {Papa}, {Eggenstein}, {Allen},
  {Dergachev}, {Prix}, {Machenschalk}, {Walsh}, {Zhu}, \&
  {Kwang}}]{2020arXiv200912260S}
{Steltner}, B., {Papa}, M.~A., {Eggenstein}, H.~B., {et~al.} 2020, arXiv
  e-prints, arXiv:2009.12260

\bibitem[{Woan {et~al.}(2018)Woan, Pitkin, Haskell, Jones, \&
  Lasky}]{Woan_2018}
Woan, G., Pitkin, M.~D., Haskell, B., Jones, D.~I., \& Lasky, P.~D. 2018, The
  Astrophysical Journal, 863, L40

\bibitem[{Zimmermann \& Szedenits(1979)}]{PhysRevD.20.351}
Zimmermann, M. \& Szedenits, E. 1979, Phys. Rev. D, 20, 351

\end{thebibliography}

\end{document}